\documentclass[pdflatex,sn-mathphys-num]{sn-jnl}
\usepackage{graphicx}%
\usepackage{multirow}%
\usepackage{amsmath,amssymb,amsfonts}%
\usepackage{amsthm}%
\usepackage{amsmath}
\usepackage{mathrsfs}%
\usepackage[title]{appendix}%
\usepackage{xcolor}%
\usepackage{textcomp}%
\usepackage{manyfoot}%
\usepackage{booktabs}%
\usepackage{algorithm}%
\usepackage{algorithmicx}%
\usepackage{algpseudocode}%
\usepackage{listings}%
\theoremstyle{thmstyleone}

\theoremstyle{thmstyletwo}%
\theoremstyle{thmstylethree}%

\begin{document}
\title[Article]{Field-free dual superconducting diode via photon-assisted interference}

\author*[1]{\fnm{Laura} \sur{Borgongino}}\email{laura.borgongino@sns.it}
\author*[2,3]{\fnm{Rubén} \sur{Seoane Souto}}\email{ruben.seoane@csic.es}
\author[2,3]{\fnm{José Luis} \sur{del Olmo}}
\author[2,3]{\fnm{Ramón} \sur{Aguado}}
\author[1]{\fnm{Alessandro} \sur{Crippa}}
\author[1]{\fnm{Francesco} \sur{Giazotto}}
\author*[1]{\fnm{Elia} \sur{Strambini}}\email{elia.strambini@cnr.it}

\affil[1]{Istituto Nanoscienze-CNR, NEST-SNS, Piazza San Silvestro 12, 56127 Pisa, Italy}
\affil[2]{Instituto de Ciencia de Materiales de Madrid (ICMM), Consejo Superior de Investigaciones Científicas (CSIC), Sor Juana Inés de la Cruz 3, 28049 Madrid, Spain}
\affil[3]{Quantum Advanced Research Center (QuARC), Consejo Superior de Investigaciones Científicas (CSIC), Sor Juana Inés de la Cruz 3, 28049 Madrid, Spain}

\abstract{Non-reciprocal transport underpins key functionalities in signal processing and logic; however, conventional semiconductor diodes exhibit markedly reduced performance at sub-Kelvin temperatures. As cryogenic electronics and quantum technologies continue to scale, there is increasing demand for diode operation compatible with low-temperature and low-dissipation conditions~\cite{nadeem_quantum_2026}. This need has driven the development of alternatives to p--n junctions that rely on non-reciprocal electron transport mechanisms efficient in the deep-cryogenic regime.
Superconducting devices constitute a natural low-loss platform, yet most existing superconducting diodes are restricted to either Cooper-pair\cite{ando_observation_2020,borgongino_biharmonic-drive_2025} or quasiparticle transport~\cite{strambini_superconducting_2022}. Consequently, they often require complex material stacks, deliberately asymmetric device geometries, or external magnetic fields. No single, geometrically symmetric junction has yet integrated rectification across both transport channels to achieve vanishing resistance under forward bias alongside strongly suppressed conduction under reverse bias.
Here, we demonstrate a dynamically reconfigurable dual-function superconducting diode based on a conventional $\mathrm{Al/AlO_x/Al}$ tunnel junction. By breaking spatiotemporal symmetry via microwave biharmonic driving, we exploit multi-tone photon-assisted tunneling to independently control dissipationless (Cooper-pair) and dissipative (quasiparticle) transport channels within the same device. 
Varying the microwave drive amplitude induces a transition from a supercurrent diode to an ideal quasiparticle diode, achieving rectification efficiencies exceeding those of conventional Schottky diodes and contemporary superconducting diodes, while the relative phase between harmonics deterministically reverses the diode polarity. We further demonstrate AC signal rectification under both current-bias and voltage-bias configurations. Finally, we implement a proof-of-concept “absolute diode” that combines zero forward resistance with strongly suppressed reverse conduction. This tunable approach provides a versatile building block for low-power cryogenic electronic architectures.}


\maketitle

\begin{figure}[h]
\centering
\includegraphics[width=0.9\textwidth]{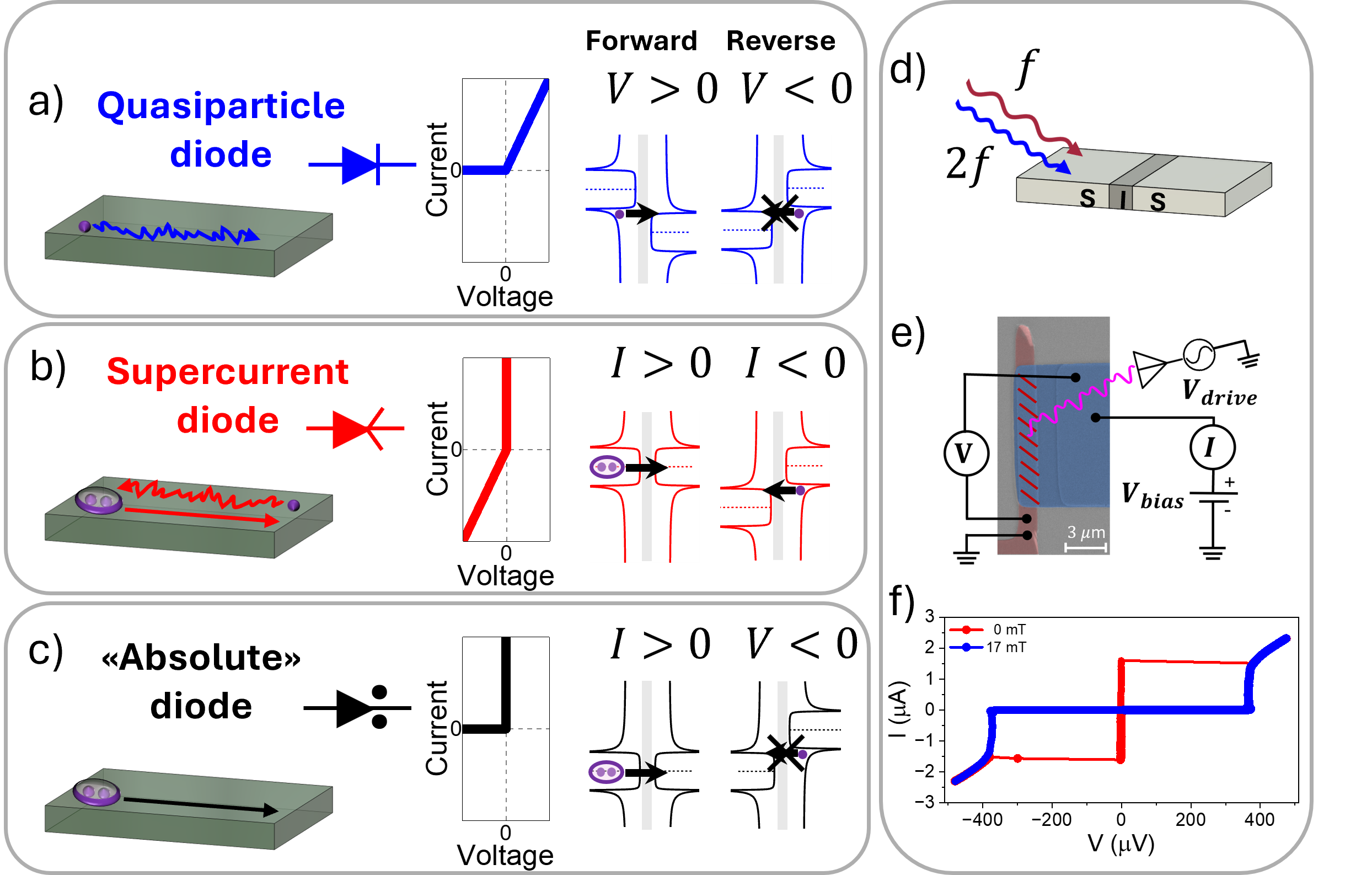}
\caption{\textbf{Operational concepts of superconducting diodes, the absolute diode paradigm, and device implementation.} 
\textbf{a}, \textbf{Quasiparticle diode.} Conceptual sketch of quasiparticle transport, circuit symbol, and $I\text{--}V$ curve under voltage bias. Forward conduction occurs ($R_{\mathrm{FWD}} > 0$) alongside suppressed reverse conduction ($R_{\mathrm{REV}} \rightarrow \infty$), analogously to a semiconductor p--n junction. 
\textbf{b}, \textbf{Supercurrent diode.} Conceptual sketch of Cooper-pair (forward) and quasiparticle (backward) transport mechanisms, circuit symbol, and current-biased $I\text{--}V$ characteristic. Dissipationless transport ($R_{\mathrm{FWD}} = 0$) converts to dissipative conduction ($R_{\mathrm{REV}} > 0$) under reverse bias. 
\textbf{c}, \textbf{Absolute diode.} Conceptual sketch of combined carrier tunneling, circuit symbol, and ideal $I\text{--}V$ characteristic. The device synthesizes zero forward resistance ($R_{\mathrm{FWD}} = 0$, via Cooper pairs) with infinite reverse resistance ($R_{\mathrm{REV}} \rightarrow \infty$, via quasiparticle suppression). 
\textbf{d}, Schematic layout of an SIS tunnel junction irradiated by a biharmonic microwave drive composed of frequencies $f$ and $2f$.
\textbf{e}, False-color scanning electron micrograph (SEM) of the $\mathrm{Al/AlO_x/Al}$ junction and four-wire experimental setup.  
\textbf{f}, Undriven $I\text{--}V$ traces recorded at zero field ($B = 0\,\mathrm{mT}$, red) and a Fraunhofer pattern minimum  ($B = 17\,\mathrm{mT}$, blue). 
At $B = 0\,\mathrm{mT}$, the junction exhibits a symmetric switching current $I_{\mathrm{sw}}^{+} = |I_{\mathrm{sw}}^{-}| \approx 1.6\,\mu\mathrm{A}$ and a normal-state resistance $R_{\mathrm{N}} \approx 160\,\Omega$.}
\label{fig1}
\end{figure}

Superconducting platforms offer a promising framework enabling cryogenic non-reciprocal transport in both dissipative quasiparticle and dissipationless Cooper-pair channels, despite the native electron-hole symmetry of conventional superconductors.

Quasiparticle superconducting diodes exhibit a unidirectional suppressed-current region associated with the superconducting gap and have been engineered with spin filtering\cite{strambini_superconducting_2022} or subgap features such as Yu-Shiba-Rusinov resonances\cite{trahms_diode_2023}. Similar to conventional p--n diodes, an ideal quasiparticle diode exhibits a finite resistance ($R_\mathrm{FWD}>0$) and an exponentially large resistance ($R_\mathrm{REV}\rightarrow \infty$) in the forward and backward \emph{Voltage} bias, respectively (Figure \ref{fig1}a).

Conversely, Cooper-pair supercurrent diodes\cite{ando_observation_2020,he_phenomenological_2022,zhang_general_2022,turini2022josephson,ilic_theory_2022,bauriedl_supercurrent_2022,narita_field-free_2022,wu_field-free_2022,souto_josephson_2022,pal_josephson_2022,margineda2023sign,yun_magnetic_2023,chiles_nonreciprocal_2023,gupta_gate-tunable_2023,paolucci2023gate,ghosh_high-temperature_2024,li_unconventional_2024,scheer_tunable_2025,orban_superconducting_2026} 
operate under current bias, supporting a dissipationless state under forward bias ($R_{\mathrm{FWD}} = 0$) up to the critical current threshold and a dissipative conducting state in the reverse direction ($R_\mathrm{REV}>0$), thereby complementing classical p--n and quasiparticle diodes (Fig.~\ref{fig1}b).

Synthesizing both mechanisms inside a single junction leads to the ultimate concept of an \emph{absolute diode},  offering zero forward resistance ($R_{\mathrm{FWD}} = 0$) and infinite reverse resistance ($R_{\mathrm{REV}} \rightarrow \infty$) then eliminating dissipative losses in the off-state while guaranteeing zero power dissipation in the on-state (Figure \ref{fig1}c). 
Among superconducting devices, the simple superconductor-insulator-superconductor (SIS) tunnel junction provides a prototypical platform for this diode because it naturally hosts two distinct transport channels: coherent Cooper-pair tunneling and dissipative quasiparticle tunneling with gap suppression.

Here, we demonstrate that a conventional, symmetric SIS tunnel junction under biharmonic microwave excitation operates as a dynamically reconfigurable dual superconducting diode for both Cooper pairs and quasiparticles.
By leveraging multi-tone photon-assisted tunneling, we break temporal inversion symmetry electrodynamically, avoiding the need for static magnetic fields or structural material asymmetries~\cite{borgongino_biharmonic-drive_2025}.

\section*{Diode implementation and Operating Mechanism}\label{sec2}

The operating principle of the dual superconducting diode relies on dynamic symmetry breaking via multi-tone photon-assisted tunneling in an SIS junction (Fig.~\ref{fig1}d). Absorption or emission of $n$ photons of frequency $f$ shifts the transition energies by $\Delta E = \pm n h f$. For Cooper pairs (charge $2e$), this assistance forms Shapiro steps at $V_{\mathrm{Shapiro}} = n h f / 2e$~\cite{shapiro_josephson_1963}. For single quasiparticles (charge $e$), it drives Tien-Gordon subgap transport shifted by $\Delta V_{\mathrm{QP}} = n h f / e$~\cite{dayem_quantum_1962,tien_multiphoton_1963,tinkham_introduction_2004,platero_photon-assisted_2004}.

Monochromatic driving preserves the time-averaged inversion symmetry, producing reciprocal $I\text{--}V$ curves. To induce non-reciprocity, we apply a biharmonic excitation combining a fundamental frequency $f$ and its second harmonic $2f$:
\begin{equation}
V_{\mathrm{drive}}(t) = V_1 \sin(2\pi f t) + V_2 \sin(4\pi f t + \theta).
\label{eq:drive}
\end{equation}
The tuning of the relative phase $\theta$ modifies the temporal symmetry of the waveform, reaching the maximum breaking of the symmetry at $\theta = \pm \pi/2$~\cite{scheer_tunable_2025}.
In the Josephson regime, this wave-shape asymmetry promotes non-reciprocal multi-photon exchange during Cooper-pair tunneling, resulting in a tunable supercurrent diode effect~\cite{borgongino_biharmonic-drive_2025}. Simultaneously, it produces asymmetric multi-photon absorption and emission for single quasiparticles, unifying Cooper-pair and single-electron rectification in one system.

We implement this concept in an $\mathrm{Al/AlO_x/Al}$ junction made by shadow-mask evaporation (Fig.~\ref{fig1}e). In the undriven state at zero magnetic field, the junction exhibits a symmetric switching current $I_{\mathrm{sw}}^{+} = |I_{\mathrm{sw}}^{-}| \approx 1.6\,\mu\mathrm{A}$, normal-state resistance $R_{\mathrm{N}} \approx 160\,\Omega$, and a retrapping current $I_{\mathrm{r}} \approx 50\,\mathrm{nA}$ (Fig.~\ref{fig1}f, red curve, and Extended Fig. \ref{figED1}a). Suppressing the supercurrent at a Fraunhofer magnetic field minimum ($B = 17\,\mathrm{mT}$, extended Fig. \ref{figED1}b) isolates quasiparticle transport, revealing a flat subgap region for $|V| < 2\Delta/e$ with gap $\Delta \approx 189\,\mu\mathrm{eV}$ and subgap resistance $R_{\mathrm{subgap}} \approx 0.5\,\mathrm{M}\Omega$ (Fig.~\ref{fig1}f, blue curve). More information about the sample can be found in Supplementary Information Section 1.

\section*{Quasiparticle biharmonic tunnel diode}\label{sec3}

\begin{figure}[!htbp]
\centering
\includegraphics[width=0.9\textwidth]{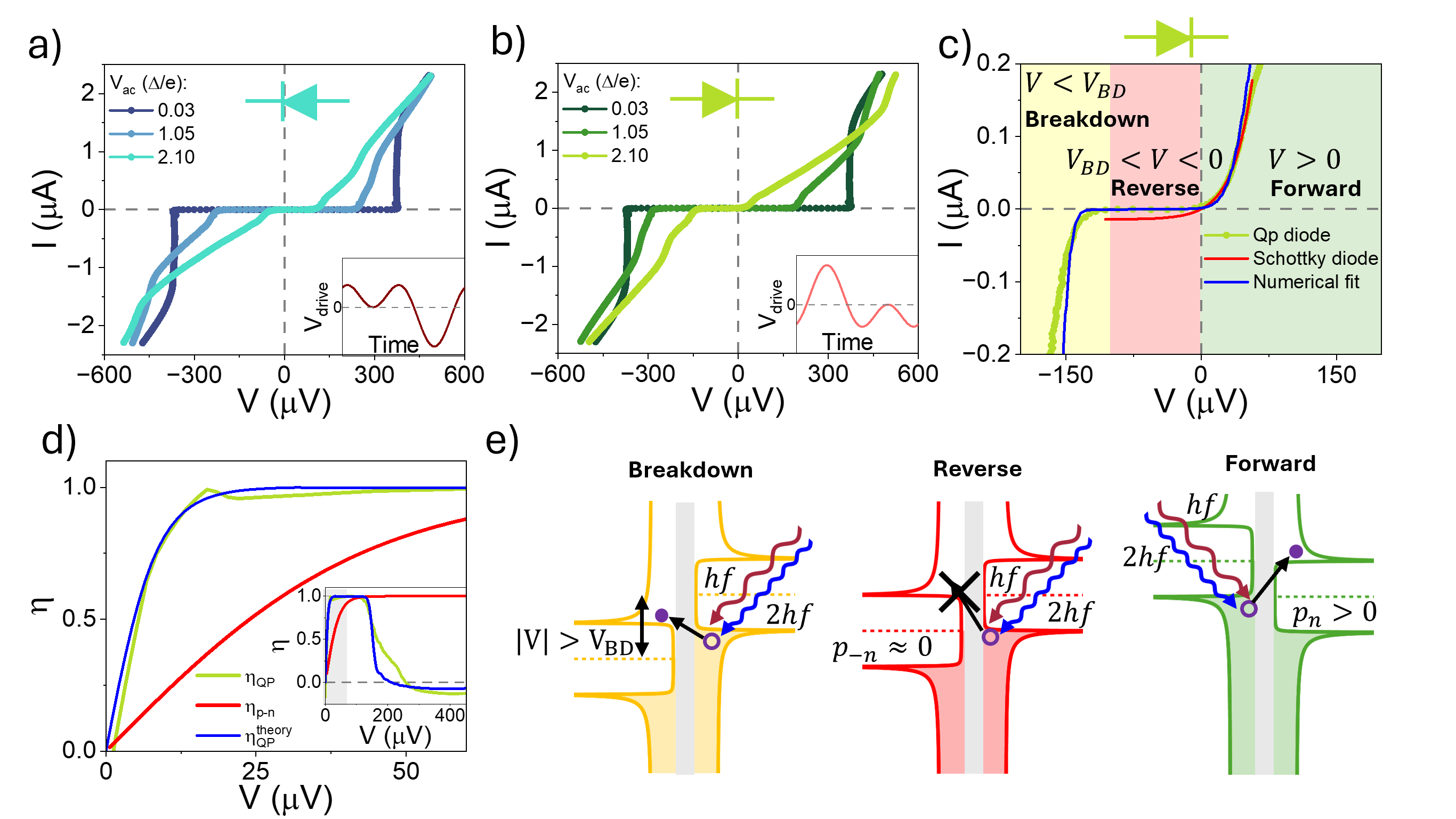}

\caption{\textbf{Biharmonic drive-induced quasiparticle diode.}
\textbf{a}, \textbf{b}, $I\text{--}V$ characteristics under biharmonic driving  ($f = 1.405\,\mathrm{GHz}$ and $2f = 2.810\,\mathrm{GHz}$) at various drive amplitudes $V_{\mathrm{ac}}$ for relative phases  $\theta = \pi/2$ (\textbf{a}) and $\theta = -\pi/2$ (\textbf{b}). Insets: time-domain waveforms of the biharmonic drive $V_{\mathrm{drive}}(t)$. 
\textbf{c}, Ideal quasiparticle diode trace ($\theta = -\pi/2$, green) compared with a semiconductor Schottky diode fit (red, $n = 3.1$, $I_{\mathrm{S}} = 14.2\,\mathrm{nA}$) and extended Tien-Gordon model (blue).
\textbf{d}, Quasiparticle diode efficiency $\eta_{\mathrm{QP}}$ (green), compared with the theoretical semiconductor p--n diode efficiency $\eta_{\mathrm{p\text{-}n}}$ (red), and the Tien-Gordon model (blue). Inset: Extended voltage range showing reverse breakdown.  
\textbf{e}, Energy-diagram schematics illustrating photon-assisted quasiparticle tunneling pathways under forward, reverse, and breakdown bias.
All data were recorded at $T \approx 80\,\mathrm{mK}$ and at a Fraunhofer pattern minimum ($B = 17\,\mathrm{mT}$) to fully suppress the supercurrent and isolate single-quasiparticle transport; zero-field measurements yield similar behavior (Supplementary Information Section 2). Drive amplitudes $V_{\mathrm{ac}}=V_1 = 2V_2$ are normalized to the superconducting gap ($\Delta \approx 189\,\mu\mathrm{eV}$). 
}
\label{fig2}
\end{figure}

Irradiating the junction at a Fraunhofer minimum ($B = 17\,\mathrm{mT}$) with a biharmonic drive of frequency $f = 1.405\,\mathrm{GHz}$ and $2f =2.81\,\mathrm{GHz}$ and relative phase difference $\theta = \pi/2$ transforms quasiparticle transport into a non-reciprocal diode state as the microwave amplitude increases (Fig.~\ref{fig2}a).
We calibrated the effective AC voltage drop across the junction using standard monochromatic photon-assisted tunneling measurements (Supplementary Information Section 3). To maximize waveform asymmetry,  the amplitudes of the two harmonics were fixed at $V_1 = 2V_2 = V_{\mathrm{ac}}$ (Supplementary Information Section 4).

At drive amplitudes that match the gap scale ($\max[|V_{\mathrm{drive}}(t)|] = 2\Delta/e$), ideal diode efficiency is achieved with forward current flowing freely while the reverse current is suppressed over a finite voltage window. Inverting the phase to $\theta = -\pi/2$ reverses the diode polarity (Fig.~\ref{fig2}b), confirming that the diode polarity is dynamically dictated by the spatio-temporal asymmetry of the driving waveform.

At $\theta = -\pi/2$, a close inspection of the $I\text{--}V$ trace reveals a forward current that increases exponentially, matching the characteristic of the Schottky diode $I_{\mathrm{pn}}(V) = I_{\mathrm{S}} [ \exp(e V / n k_{\mathrm{B}} T) - 1 ]$ (Fig.~\ref{fig2}c, green and red curves). Under reverse bias, the quasiparticle diode exhibits stronger current suppression than semiconductor diodes, saturating at nearly zero reverse current and limited only by residual subgap leakage until breakdown occurs at $V_{\mathrm{BD}} = 2\Delta/e + \min[V_{\mathrm{drive}}(t)] \approx 100\,\mu\mathrm{V}$, similar to a Zener diode.

Moreover, evaluating the diode efficiency\cite{strambini_superconducting_2022}:
\begin{equation}
\eta_{\mathrm{QP}}(V) = \frac{I(V) - |I(-V)|}{I(V) + |I(-V)|}.
\label{eq:eta_qp}
\end{equation}
shows that $\eta_{\mathrm{QP}} \approx 1$ is reached at lower voltages than the semiconductor counterparts ($\eta_{\mathrm{pn}} = \tanh[e V / (2 n k_{\mathrm{B}} T)]$), confirming the superior low-temperature performance (Fig.~\ref{fig2}d).

Microscopically, extending the Tien-Gordon model to a biharmonic photon-assisted drive stems from tunneling theory evaluated via Fermi's golden rule. In this framework, single-quasiparticle transport is governed by transition rates proportional to the squared effective matrix element and the joint density of states, $W_{\mathrm{QP}} \propto \frac{2\pi}{\hbar} |M|^2 D_1(E) D_2(E + e V)$. In the undriven equilibrium state, single-electron tunneling is strictly suppressed within the subgap voltage window $|V| < 2\Delta/e$, and the symmetric density of states $D(E) = D(-E)$ yields symmetric $I\text{--}V$ characteristics.

Under monochromatic microwave driving at frequency $f$, photon absorption and emission dress the quasiparticle states, opening photon-assisted tunneling channels at energies shifted by $nhf$. Within the Tien-Gordon framework, the transition rates evaluated via Fermi's golden rule are modulated by the probability of exchanging $n$ photons
\begin{equation}
I(V,V_{\rm{ac}}) = \sum_{n=-\infty}^{+\infty} p_n(V_{\rm{ac}}) I_0(V + nhf/e),
\label{eq:I_V_1harmonic}
\end{equation}
where $I_0(V)$ is the non-perturbed quasiparticle current and $p_n = J_n^2(e V_{\mathrm{ac}} / h f)$ represents the probability of exchanging $n$ photons of frequency $f$ and $J_n$ is the $n$-th order Bessel function of the first kind.
Because $J_n^2 = J_{-n}^2$, photon absorption and emission transition rates remain intrinsically identical under single-tone driving. Combined with the anti-symmetry of the unperturbed state $I_0(-V) = -I_0(V)$, this symmetry preserves transport reciprocity under bias reversal  $I(-V, V_{\mathrm{ac}}) = -I(V, V_{\mathrm{ac}})$.

Under multi-tone biharmonic driving with frequencies $f$ and $2f$, quantum interference between multi-photon transition pathways fundamentally alters this picture. Generalized multi-photon amplitudes now govern the transition probabilities. This results in phase-dependent exchange probabilities:
\begin{equation}
p_n(V_1,V_2,\theta)=\left|\sum_{m=-\infty}^\infty J_{n-2m}(eV_1/hf)J_{m}(eV_2/2hf)e^{im(\theta-\pi/2)}\right|^2.
\label{eq:I_V_2harmonics}
\end{equation}
For a net energy transfer $nhf$, this sum runs over all coherent combinations of photon absorption and emission events at $f$ and $2f$. Crucially, the relative drive phase $\theta$ determines whether these microscopic transition amplitudes interfere constructively or destructively. For $\theta \neq 0, \pi$ (mod $\pi$), this multi-path interference breaks the symmetry between forward and backward transition rates ($p_n \neq p_{-n}$), unbalancing Fermi's golden rule transition probabilities for opposite bias polarities. This microscopic asymmetry selectively promotes single-quasiparticle tunneling in one bias direction while suppressing it in the reverse direction, realizing a dynamically tunable quasiparticle diode.

The numerical results obtained from this extended Tien-Gordon framework (Fig.~\ref{fig2}c,d, blue curves) show remarkable agreement with the experimental data across all operating regimes of the quasiparticle diode.
Further details on the analytical derivation of $p_n$ and numerical simulations are provided in Supplementary Information Section 5 and 6.

\section*{Superconducting biharmonic tunnel diode}\label{sec4}

\begin{figure}[h]
\centering
\includegraphics[width=0.9\textwidth]{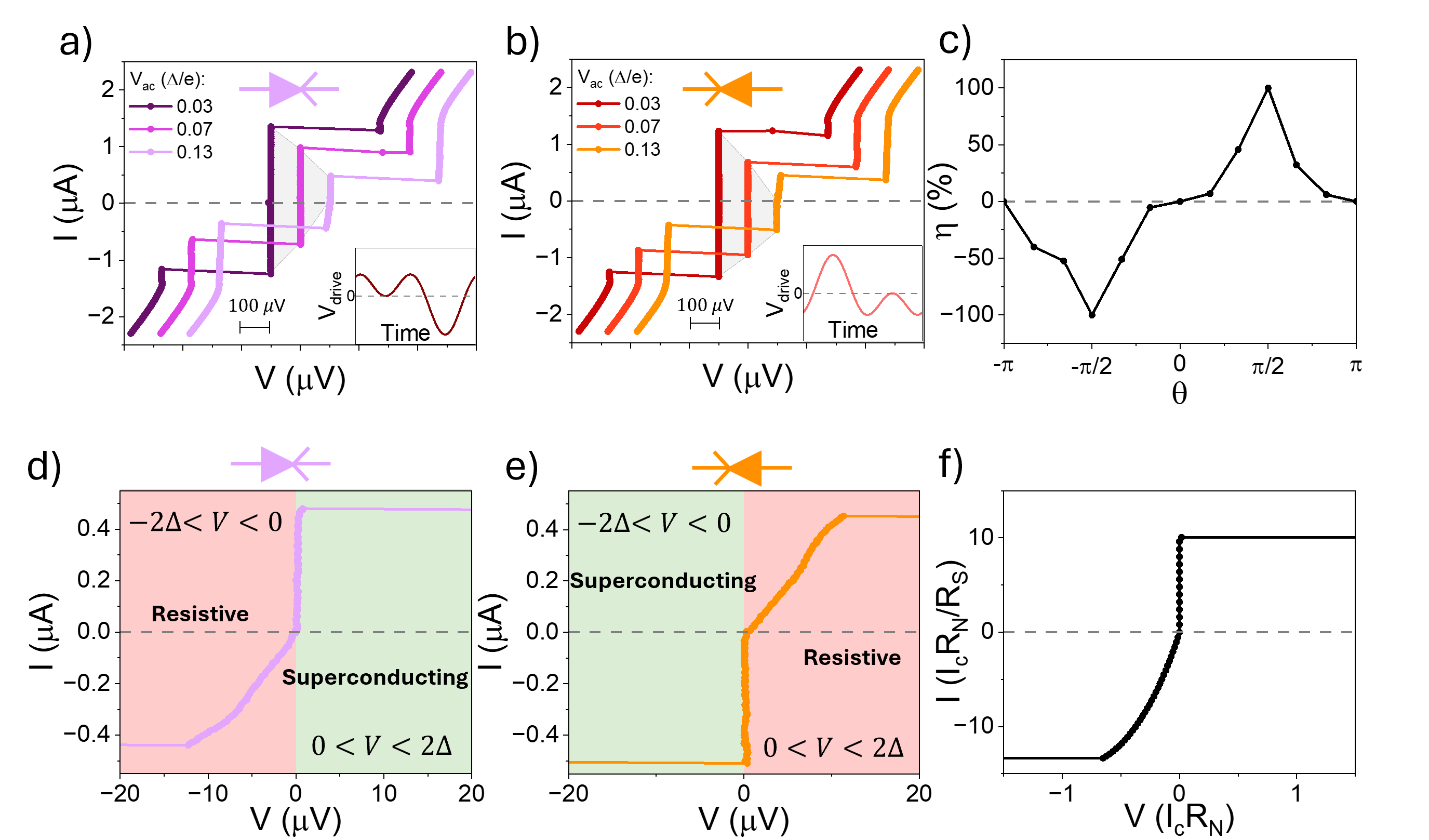}
\caption{\textbf{Drive-induced supercurrent diode and polarity inversion.} 
\textbf{a}, \textbf{b}, $I\text{--}V$ characteristics recorded under biharmonic microwave driving ($f = 1.405\,\mathrm{GHz}$, $2f = 2.810\,\mathrm{GHz}$) at increasing amplitude $V_{\mathrm{ac}}$ for relative phases $\theta = \pi/2$ (\textbf{a}) and $\theta = -\pi/2$ (\textbf{b}).
\textbf{c}, Supercurrent diode efficiency $\eta_{\mathrm{S}}$ versus relative phase $\theta$ at $V_{\mathrm{ac}} \approx 0.12\,\Delta/e$, displaying sinusoidal control and polarity reversal.
\textbf{d}, \textbf{e}, High-resolution $I\text{--}V$ traces in the ideal diode regime ($\eta_{\mathrm{S}} \approx 1$) for $\theta = \pi/2$ (\textbf{d}) and $\theta = -\pi/2$ (\textbf{e}), showing a finite-slope dissipative branch ($R \approx 0.15\,R_{\mathrm{N}}$) prior to switching into the subgap quasiparticle state. 
\textbf{f}, Simulated  $I\text{--}V$ characteristics obtained from the modified RSJ model for $\omega=0.001$, $V_1=2V_2=R_\mathrm{S}I_\mathrm{c}/3$, $\theta=\pi/2$, $R_\mathrm{N}=1$, and $R_\mathrm{S}=0.05$ (with $I_{\mathrm{c}}$ taken as the unit of current).
All experimental data were acquired at $T \approx 80\,\mathrm{mK}$.}
\label{fig3}
\end{figure}

At zero magnetic field, non-reciprocal Cooper-pair transport emerges with the same drive signal but at lower powers ($e V_{\mathrm{ac}} \approx 0.03\text{--}0.13\,\Delta$) than required for quasiparticles. Increasing $V_{\mathrm{ac}}$ produces switching current asymmetry ($I_{\mathrm{sw}}^{+} \neq |I_{\mathrm{sw}}^{-}|$, Fig.~\ref{fig3}a,b). The supercurrent diode efficiency:
\begin{equation}
\eta_{\mathrm{S}} = \frac{I_{\mathrm{sw}}^{+} - |I_{\mathrm{sw}}^{-}|}{I_{\mathrm{sw}}^{+} + |I_{\mathrm{sw}}^{-}|}.
\label{eq:eta_supercurrent}
\end{equation}
modulates sinusoidally with phase $\theta$, reaching $\eta_{\mathrm{S}} \approx 1$ at $V_{\mathrm{ac}} \approx 0.13\,\Delta/e$ (Fig.~\ref{fig3}c). Inverting the phase from $\theta = \pi/2$ to $\theta = -\pi/2$ dynamically reverses the diode polarity.

Notably, the supercurrent diode exhibits an opposite rectification polarity in comparison to the quasiparticle diode at identical phase settings. This polarity inversion highlights complementary underlying physics: phase-controlled interference that promotes single-quasiparticle tunneling under forward bias simultaneously modulates multi-photon phase locking for Cooper pairs in reverse bias. 

Moreover, high-resolution $I\text{--}V$ traces in the ideal diode state ($\eta_{\mathrm{S}} \approx 1$, Fig.~\ref{fig3}d,e) reveal a unique feature of the biharmonic SIS tunnel diode compared to its SNS counterpart\cite{borgongino_biharmonic-drive_2025}. 
For suppressed bias polarity, an intermediate dissipative branch ($R \approx 0.15\,R_{\mathrm{N}}$) precedes the full switching to the dissipative-voltage state (Fig.~\ref{fig3}d,e). This feature originates from photon-assisted phase-slip dynamics, as validated by RSJ circuit simulations (Fig.~\ref{fig3}f, Extended Data Fig. \ref{figED2}, and Supplementary Information Section 7).

\section*{AC rectification and Absolute diode regimes}\label{sec5}

\begin{figure}[h!]
\centering
\includegraphics[width=0.9\textwidth]{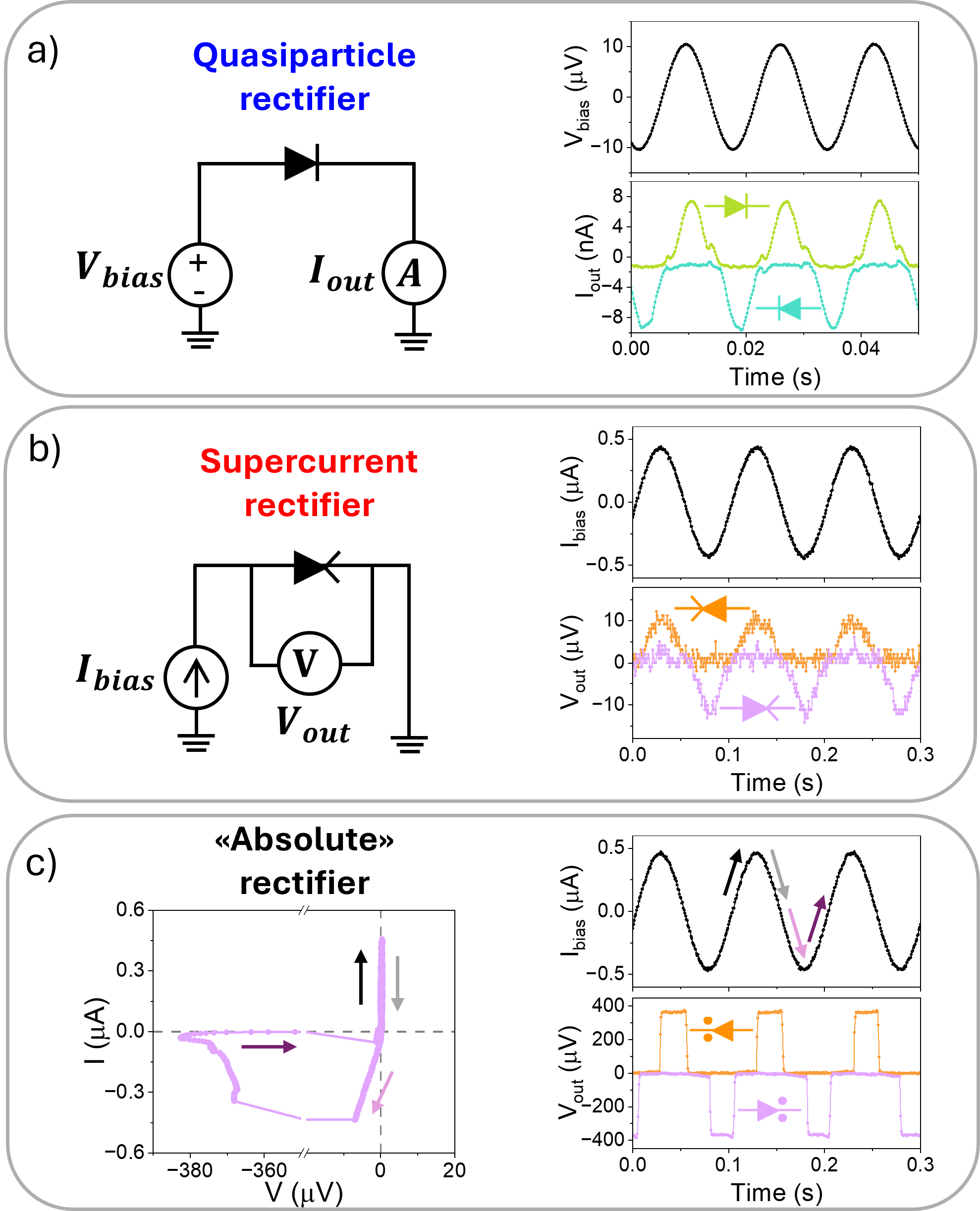}
\caption{\textbf{Low-frequency AC rectification across operating regimes.} 
\textbf{a}, Half-wave current rectification in the quasiparticle diode of an AC voltage bias ($f_{\mathrm{bias}} = 61\,\mathrm{Hz}$) for relative phases $\theta = \pm \pi/2$.
\textbf{b}, Half-wave voltage rectification in the supercurrent diode regime of an AC current bias ($f_{\mathrm{bias}} = 10\,\mathrm{Hz}$) for $\theta = \pm \pi/2$. 
\textbf{c}, Left: Static $I\text{--}V$ characteristic ($\theta = \pi/2$) highlighting the pronounced hysteresis between switching ($I_{\mathrm{sw}} \approx 450\,\mathrm{nA}$) and retrapping ($I_{\mathrm{r}} \approx 50\,\mathrm{nA}$) currents. Arrows denote the bias sweep direction. Right: Square-wave voltage rectification recorded under an AC current bias ($f_{\mathrm{bias}} = 10\,\mathrm{Hz}$), clipped strictly at $V \approx 2\Delta/e$. 
All measurements were performed at $T \approx 100\,\mathrm{mK}$ under biharmonic driving ($f = 1.405\,\mathrm{GHz}$, $2f = 2.810\,\mathrm{GHz}$ and $\theta = \pm \pi/2$).
}
\label{fig4}
\end{figure}

We validate low-frequency AC signal rectification ($f_{\mathrm{bias}} \ll f_{\mathrm{drive}}$) across all operating modes accessible at different biharmonic drive amplitudes. 

In the quasiparticle diode regime, a sinusoidal voltage bias ($V^{\mathrm{AC}}_{\mathrm{bias}} < \Delta/e$) applied alongside the biharmonic drive at $\theta = \pi/2$ yields clean half-wave current rectification. 
The output current remains within the noise floor during the suppressed half-cycle and tracks the static onset during the conducting half-cycle, inverting the polarity at $\theta = -\pi/2$. 
Equivalent half-wave rectification was also recorded at different $f_{\mathrm{bias}}$ (Extended Data Fig. \ref{figED3}) and across different devices (Extended Data Fig. \ref{figED4}), confirming frequency independence over low-bias bandwidth and reproducibility over different samples.

In the supercurrent regime under current bias ($I_{\mathrm{AC}} < I_{\mathrm{sw}}^{+}$, Fig.~\ref{fig4}b), the junction produces half-wave voltage rectification with phase-tunable polarity. 

At higher AC current amplitudes, pronounced hysteresis between switching ($I_{\mathrm{sw}} \approx 450\,\mathrm{nA}$) and retrapping currents ($I_{\mathrm{r}} \approx 50\,\mathrm{nA} \approx 0.11\,I_{\mathrm{sw}}$) drives the system into an absolute diode regime (Fig.~\ref{fig4}c). 
Forward bias maintains a zero-resistance supercurrent state ($V = 0$). In reverse bias, switching to the subgap branch combined with $I_{\mathrm{r}} \ll I_{\mathrm{sw}}$ locks the junction in the dissipative state throughout the return half-cycle. 

Consequently, the output voltage saturates near  $V \approx 2\Delta/e$, converting sinusoidal inputs into flat-topped square waves. This response resolves the voltage-divergence constraint inherent to current-driven ideal rectifiers, by using switching-retrapping hysteresis to establish a stable voltage-clipped output.

\section*{Conclusions}\label{conclusions}
In summary, we have demonstrated a field-free, dynamically reconfigurable dual superconducting diode within a single SIS tunnel junction driven by a phase-controlled biharmonic microwave excitation. By leveraging multi-photon quantum interference, the relative drive phase $\theta$ breaks time-reversal symmetry electrodynamically, without requiring external magnetic fields or structural asymmetries, and simultaneously controls both dissipative quasiparticle and non-dissipative Cooper-pair transport channels.

In the quasiparticle regime, multi-photon interference yields non-reciprocal transport with vanishing reverse conductance, resembling ideal Zener-diode breakdown and outperforming conventional semiconductor Schottky diodes. 
In the Josephson regime, the device operates as an ideal supercurrent diode ($\eta_{\mathrm{S}} \approx 1$), with suppressed reverse critical current and inverted polarity relative to the quasiparticle channel. 

Furthermore, we demonstrate low-frequency half-wave AC signal rectification and establish an absolute diode operating regime under current bias, converting sinusoidal inputs into flat-topped square waves clipped at $2\Delta/e$.

Based on the standard $\mathrm{Al/AlO_x/Al}$ tunnel junction and conventional microwave control lines, this platform offers immediate compatibility with transmon qubit architectures, RSFQ digital logic, and dilution refrigerator environments. Its field-free operation eliminates flux-induced dephasing in qubit circuits, while its quantized $2\Delta/e$ voltage clipping circumvents semiconductor carrier freeze-out, offering stable logic levels for cryogenic computing, neuromorphic activation functions, and microwave energy harvesting.

Finally, the opposing rectification polarities of the quasiparticle and supercurrent channels enable their explicit experimental disentanglement, providing a direct route to systematically investigate non-equilibrium transport mechanisms and engineer directional thermoelectric functionalities.

\section*{Methods}\label{methods}
\subsection*{Sample fabrication}
The $\mathrm{Al/AlO_x/Al}$ tunnel junction was fabricated on a $\text{Si}$ substrate using double-angle shadow-mask evaporation ~\cite{dolan_offset_1977} in an ultra-high vacuum electron-beam evaporator. 
Two $30\,\mathrm{nm}$-thick Al electrodes were deposited at angles of $\pm 25^\circ$ at $1\,\mathring{\mathrm{A}}/\mathrm{s}$ to form the junction (overlap area $A \approx 16\,\mu\mathrm{m}^2$) . Controlled \textit{in-situ} thermal oxidation performed between evaporations yielded a tunnel barrier with a specific resistance $R_\square \approx 10\,\mathrm{k}\Omega\cdot\mu\mathrm{m}^2$.

\subsection*{Experimental setup}
Electrical 4-wire measurements were performed at $80\,\mathrm{mK}$ in a Leiden Cryogenics CF-CS110 dilution refrigerator equipped with a superconducting magnet
DC bias voltage, and current readout was controlled by a Keithley 2600 SMU. We amplified the differential voltages using a DL Instruments 1201 preamplifier (gain $10^3$) and recorded them  with an Agilent 34401A multimeter.
We synthesized biharmonic microwave signals using two synchronized signal generators, combined via a 2–10 GHz power splitter, attenuated at cryo-stages, and delivered via an open-ended coaxial cable placed near the sample.
For AC rectification, a function generator supplied the input signals, and the output waveforms were amplified (DL Instruments 1201 voltage amplifier, Femto current amplifier) and captured on a Tektronix TDS 6604 oscilloscope, averaged over 500 traces.

\bmhead{Data Availability}

All data supporting the findings of this study are available from the corresponding authors upon request.

\bmhead{Code Availability}

Codes are available from the corresponding authors upon request.

\bmhead{Acknowledgments}

E.S. acknowledges the Italian Project HELICS DFM.AD002.206, R.S.S, J.L.d.O, and R.A acknowledge funding from the Spanish Ministry of Science, Innovation, and Universities through AEI Grants CEX2024-001445-S (Severo Ochoa Centres of Excellence program), PID2025-169814NB-I00, PID2024-161156NB-I00 and PID2022-140552NA-I00, and the Spanish Comunidad de Madrid (CM) “Talento Program” (Project No. 2022-T1/IND-24070).

\bmhead{Author contributions}

L.B. performed the experiments and analyzed the data. E.S. and A.C. provided experimental and technical support. R.S.S., J.L.d.O., and R.A. provided theoretical support. L.B. fabricated the devices. E.S. supervised the project and conceived the experiment together with F.G.. L.B. wrote the initial draft, and all authors revised and approved the final manuscript.

\bmhead{Corresponding authors}

Correspondence and requests for materials should be addressed to L.B. and E.S. for experimental work or R.S.S. for theoretical work. 

\bmhead{Competing interests}

The authors declare no competing interests. 

\newpage

\section*{Extended Data Figures}
\begin{figure}[h]
\centering
\includegraphics[width=0.9\textwidth]{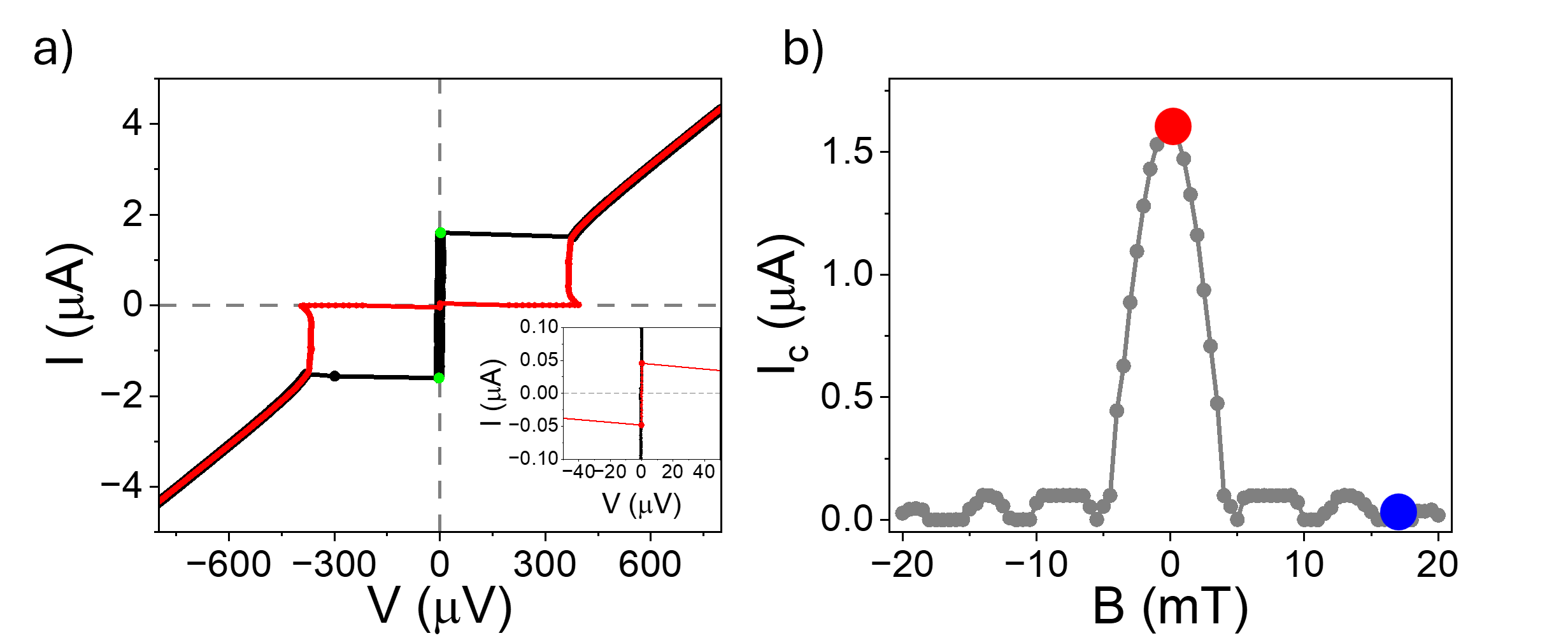}
\caption{\textbf{Full $I\text{--}V$ curve in Josephson regime and Fraunhofer interference pattern.} \textbf{a}, Extended $I\text{--}V$ curves including switching (black dots) and retrapping (red dots) branches measured at zero magnetic field. Green markers highlight equal switching currents $I_{sw}^+=|I_{sw}^-|$. Inset: Close-up of the retrapping current  showing $|I_r^\pm|\approx50,\mathrm{nA}$. 
\textbf{b}, Fraunhofer interference pattern of the switching current $I_{\mathrm{sw}}$ as a function of the in-plane magnetic field $B$. Red and blue markers highlight the magnetic fields corresponding to the maximum ($B_{\mathrm{max}} = 0\,\mathrm{mT}$) and minimum ($B_{\mathrm{min}} = 17\,\mathrm{mT}$) of the interference pattern, respectively.
Data acquired at $T \approx 80\,\mathrm{mK}$.}
\label{figED1}
\end{figure}

\newpage

\begin{figure}[h]
\centering
\includegraphics[width=0.5\textwidth]{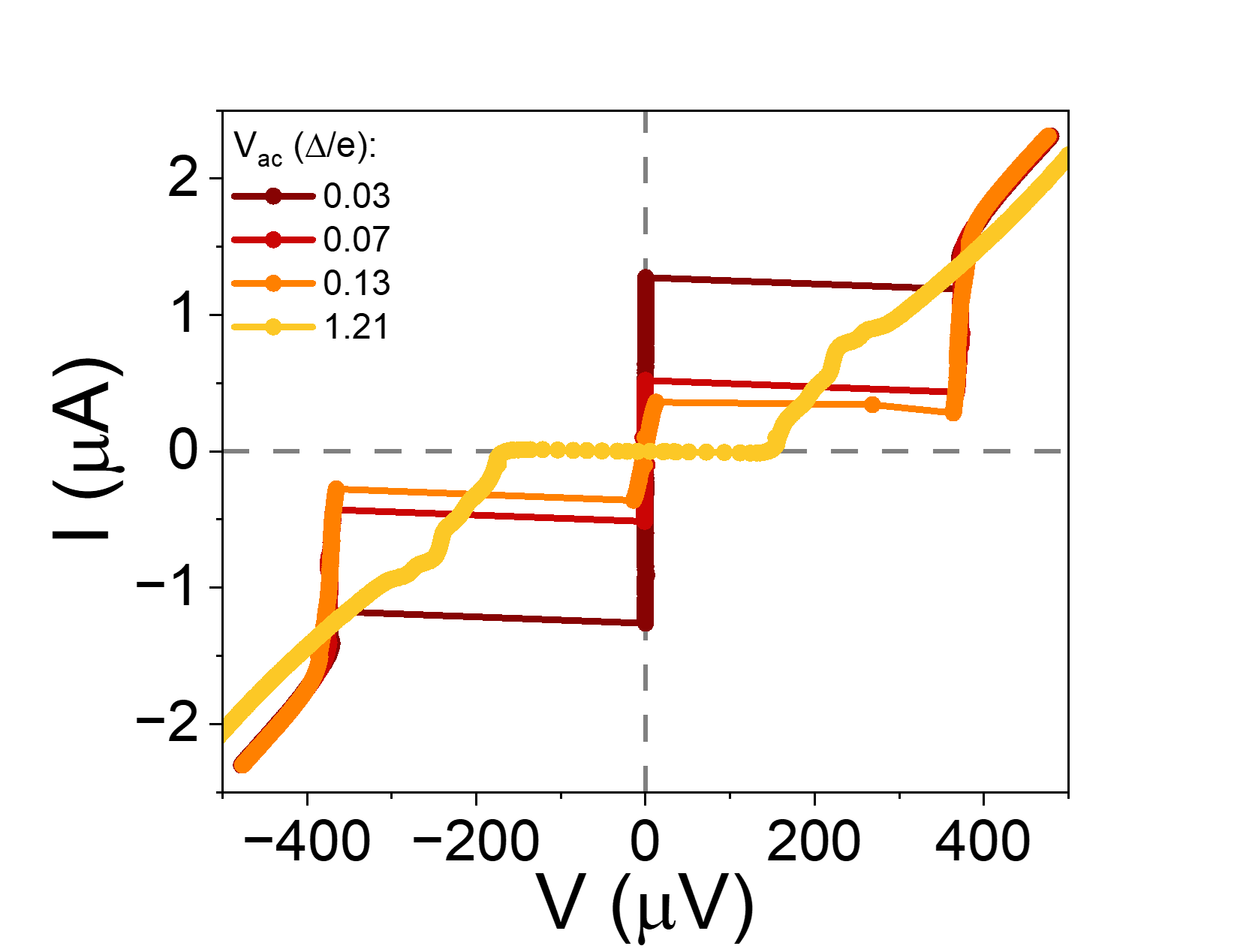}
\caption{\textbf{$I\text{--}V$ characteristics under single-tone microwave driving.} $I\text{--}V$ curves recorded under a single-tone excitation ($f=2.81 ,\mathrm{GHz}$) at various AC driving amplitudes $V_\mathrm{ac}$.}
\label{figED2}
\end{figure}

\newpage

\begin{figure}[h]
\centering
\includegraphics[width=0.5\textwidth]{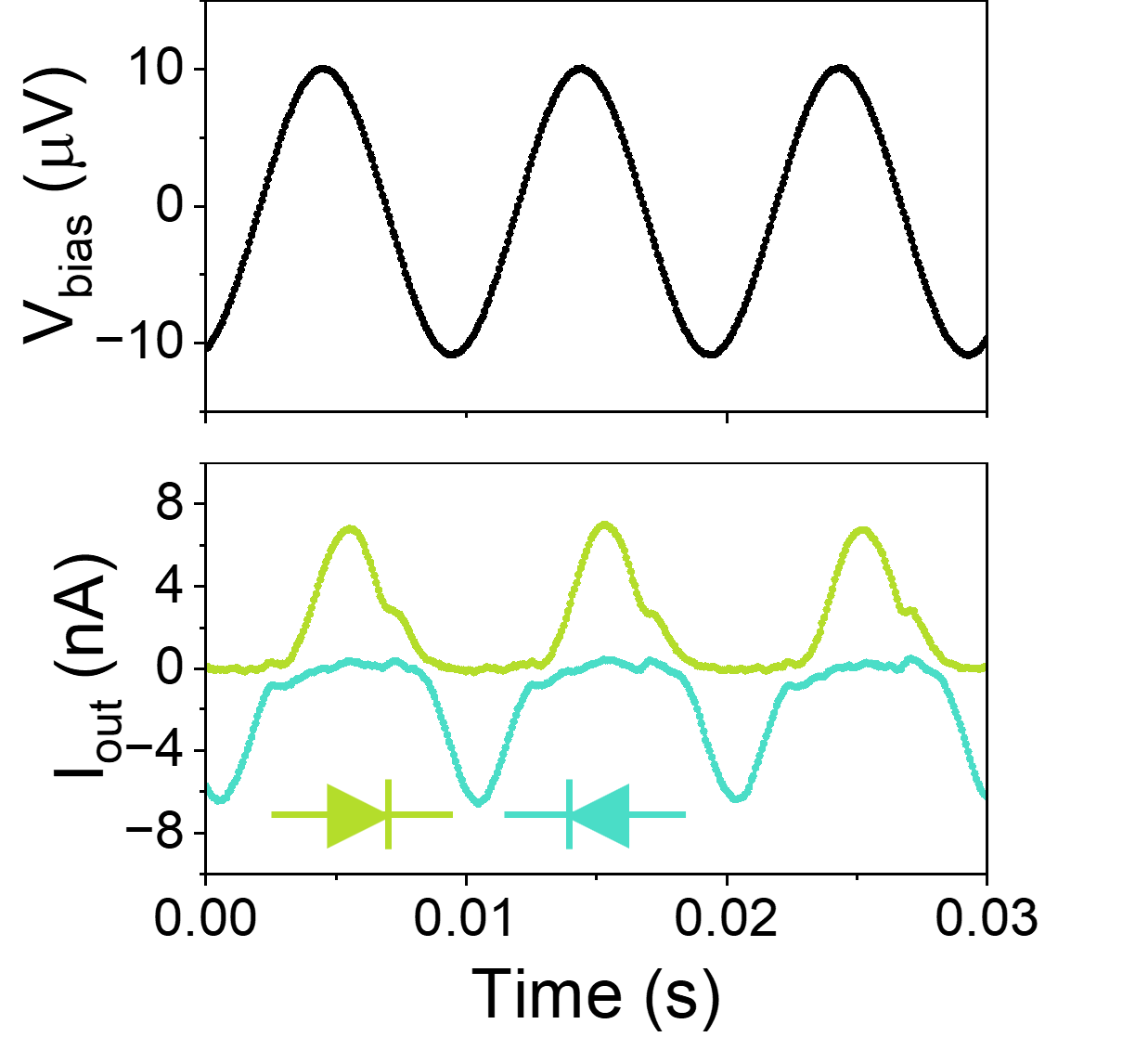}
\caption{\textbf{AC signal rectification with a quasiparticle rectifier.} 
Half-wave current rectification under AC voltage bias ($f_{\mathrm{bias}} = 101\,\mathrm{Hz}$) for relative phases $\theta = \pm \pi/2$. Measurements recorded at $T\approx100 \,\mathrm{mK}$.}
\label{figED3}
\end{figure}

\newpage

\begin{figure}[h]
\centering
\includegraphics[width=0.9\textwidth]{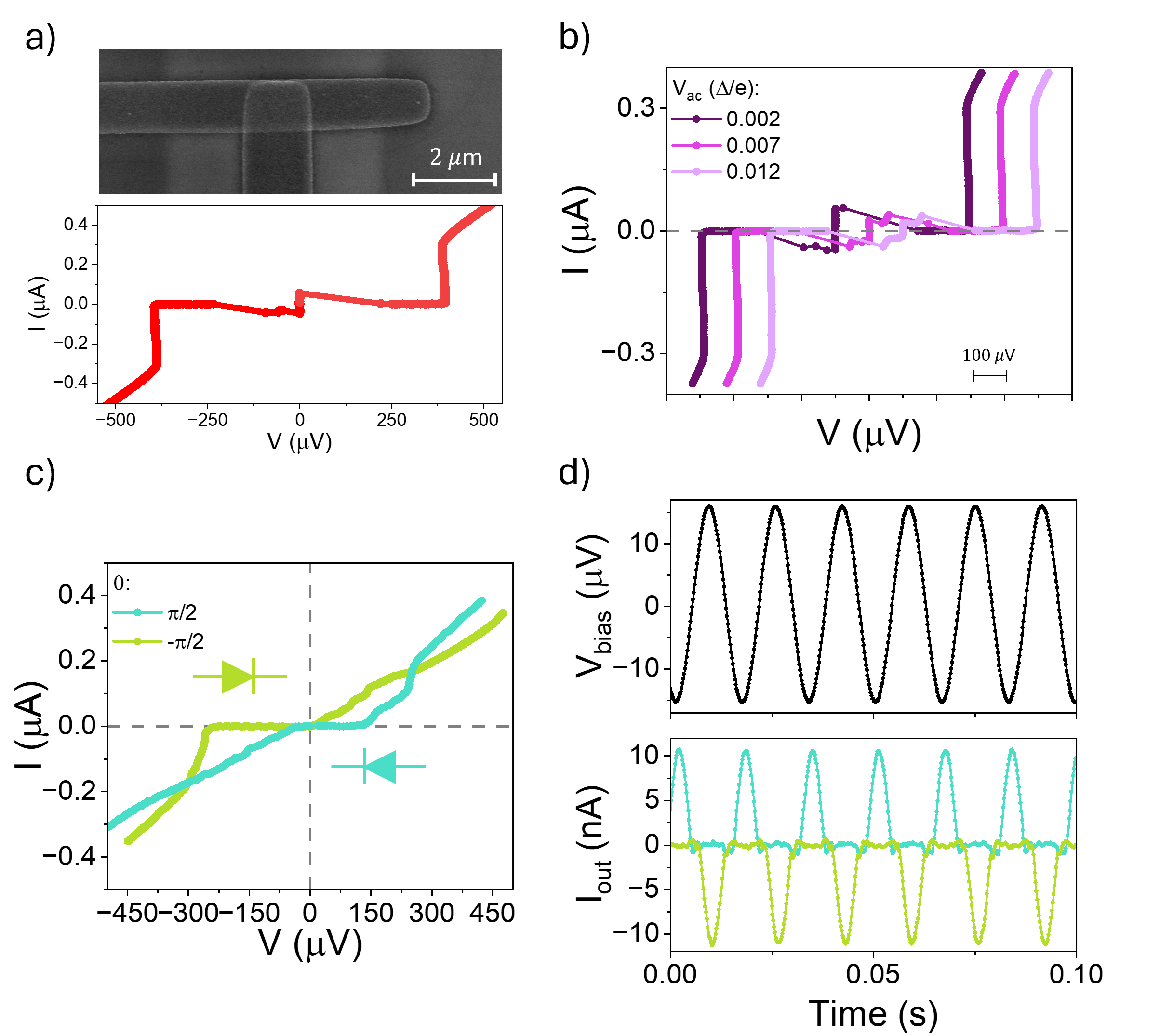}
\caption{\textbf{Quasiparticle diode characteristics in a second junction device.} 
\textbf{a}, Top: Scanning electron micrograph (SEM) of a  second device (Device 2) made with a $\mathrm{Al/AlO_x/Al}$ junction with area $A \approx 2\,\mu\mathrm{m}^2$ formed by two $30\,\mathrm{nm}$-thick aluminum leads. Bottom: $I\text{--}V$ trace recorded at a Fraunhofer maximum ($B = 0\,\mathrm{mT}$) without microwave irradiation, exhibiting $I_{\mathrm{sw}}^{+} = |I_{\mathrm{sw}}^{-}| \approx 56\,\mathrm{nA}$ and normal-state resistance $R_{\mathrm{N}} \approx 813\,\Omega$. 
\textbf{b}, $I\text{--}V$ curves under biharmonic microwave driving ($f = 1.405\,\mathrm{GHz}$, $2f = 2.810\,\mathrm{GHz}$) at increasing drive amplitudes $V_{\mathrm{ac}}$ for relative phase $\theta = \pi/2$. 
\textbf{c}, $I\text{--}V$ curves in the ideal quasiparticle diode condition $(V_{\mathrm{ac}}=2.1\Delta/e$) for harmonic phase shifts $\theta = \pi/2$ and $\theta = -\pi/2$. \textbf{d}, Half-wave current rectification under AC voltage bias ($f_{\mathrm{bias}} = 61\,\mathrm{Hz}$) for $\theta = \pm \pi/2$. }\label{figED4}
\end{figure}

\newpage
\section*{Supplementary Information}

\section{Junction parameters and hysteretic dynamics}\label{suppsec1}

The $\mathrm{Al/AlO_x/Al}$ tunnel junction has an overlap area $A = 16\,\mu\mathrm{m}^2$ and a normal-state resistance $R_{\mathrm{N}} \approx 160\,\Omega$, corresponding to a resistance-area product:
\begin{equation}
R_{\mathrm{N}} A \approx 2560\,\Omega\cdot\mu\mathrm{m}^2 = 2.56 \times 10^{-5}\,\Omega\cdot\mathrm{cm}^2.
\end{equation}
Assuming a parallel-plate capacitor model for a thin amorphous $\mathrm{AlO_x}$ with thickness $d=1\,\mathrm{nm}$ and a specific capacitance $C_s = C/A \approx 100\,\mathrm{fF}/\mu\mathrm{m}^2$\cite{deppe_determination_2004}, we estimate the total junction capacitance to be:
\begin{equation}
C = C_s \times A \approx 1.6\,\mathrm{pF}.
\end{equation}
This yields a zero-field Josephson plasma frequency of $\omega_{\mathrm{p}}/2\pi = \sqrt{e I_{\mathrm{sw}} / \pi \hbar C} \approx 23\,\mathrm{GHz}$.
The hysteretic $I\text{--}V$ characteristic observed at zero magnetic field (Extended Data Figure \ref{figED1}) is governed by the Stewart-McCumber parameter within the Resistively and Capacitively Shunted Junction (RCSJ) model:
\begin{equation}
\beta_C = \frac{2\pi}{\Phi_0} I_{\mathrm{sw}} R_{\mathrm{N}}^2 C,
\end{equation}
where $\Phi_0 = h/2e$ is the magnetic flux quantum. Substituting $I_{\mathrm{sw}} \approx 1.6\,\mu\mathrm{A}$, $R_{\mathrm{N}} = 160\,\Omega$, and $C \approx 1.6\,\mathrm{pF}$, yields $\beta_C \approx 200 \gg 1$, confirming that the junction operates in the strongly underdamped regime.

In this limit, classical RCSJ theory with an ohmic normal resistance~\cite{kautz_noise-affected_1990} predicts a retrapping current $I_{\mathrm{r}} \approx (4/\pi\sqrt{\beta_C}) I_{\mathrm{sw}} \approx 0.09 \, I_{\mathrm{sw}} \approx 144\,\mathrm{nA}$ consistent with the measured  $I_{\mathrm{r}} \approx 50\,\mathrm{nA}$. 
Key parameters of the SIS tunnel junction are summarized in Table \ref{TableS1}

\begin{table}[h!]
\centering
\caption{\textbf{Key physical and geometrical parameters of the $\mathrm{Al/AlO_x/Al}$ junction.}}\label{TableS1}
\begin{tabular}{lcc}
\hline
\textbf{Parameter} & \textbf{Symbol} & \textbf{Value} \\
\hline
Junction Area & $A$ & $16\,\mu\mathrm{m}^2$ \\
Al Lead Thickness & $t_{\mathrm{Al}}$ & $30\,\mathrm{nm}$ \\
Oxide Thickness (Estimation)& $d$ & $\approx 1\,\mathrm{nm}$\\
Normal-State Resistance & $R_{\mathrm{N}}$ & $160\,\Omega$ \\
Resistance-Area Product & $R_{\mathrm{N}} A$ & $2560\,\Omega\cdot\mu\mathrm{m}^2$ \\
Subgap Resistance& $R_{subgap}$&$0.5\text{ M}\Omega$\\
Superconducting Gap & $\Delta$ & $189\,\mu\mathrm{eV}$ \\
Dynes Parameter & $\Gamma$ & $\approx 0.02 \Delta$ \\
Switching Current ($B=0$) & $I_{\mathrm{sw}}$ & $1.6\,\mu\mathrm{A}$ \\
Retrapping Current & $I_{\mathrm{r}}$ & $50\,\mathrm{nA}$ \\
Junction Capacitance & $C$ & $\approx 1.6\,\mathrm{pF}$\\
Stewart-McCumber Parameter & $\beta_C$ & $\approx 200$\\
Plasma Frequency & $\omega_{\mathrm{p}}/2\pi$ & $\approx 23\,\mathrm{GHz}$\\
\hline
\end{tabular}
\end{table}

\section{Quasiparticle diode at zero magnetic field}\label{suppsec2}
\begin{figure}[h]
\centering
\includegraphics[width=0.9\textwidth]{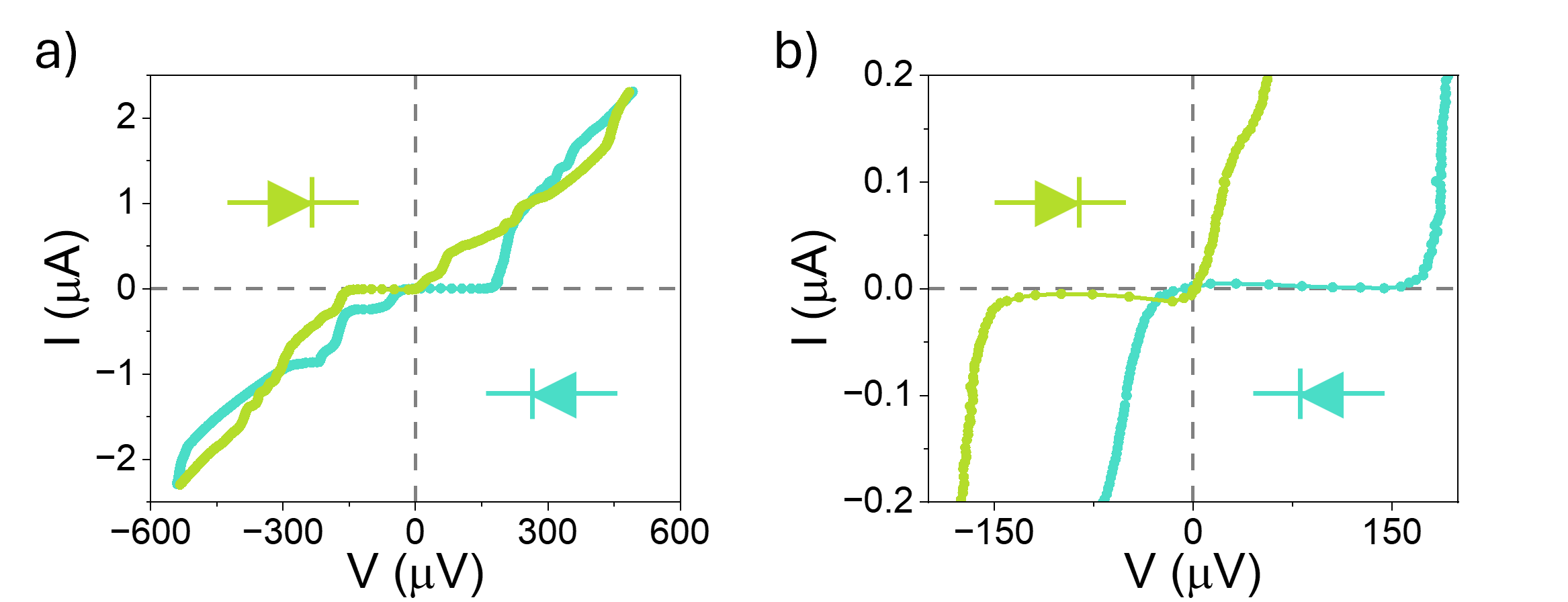}
\caption{\textbf{Quasiparticle diode with finite Josephson contribution.} 
\textbf{a},  $I\text{--}V$ characteristics recorded under biharmonic microwave excitation ($f = 1.405\,\mathrm{GHz}$ and $2f = 2.810\,\mathrm{GHz}$) in the ideal quasiparticle regime ($V_{\mathrm{ac}}=2.10\Delta/e$ for relative phases $\theta = \pi/2$ and $\theta = -\pi/2$ at zero magnetic field ($B=0\,\mathrm{mT}$). 
\textbf{b}, High-resolution view of the low-bias region. Measurements recorded at $T\approx80\,\mathrm{mK}$.}
\label{figS2}
\end{figure}

Under identical biharmonic driving condition as described in the main text  but at zero magnetic field, the device maintains its quasiparticle diode behavior while retaining a finite Josephson supercurrent contribution (Supplementary Fig.~\ref{figS2}). The $I\text{--}V$ characteristics display additional steps-like features associated with Cooper-pair photon-assisted tunneling processes, while the subgap quasiparticle rectification profile does not present relevant differences with respect to that observed at the Fraunhofer minimum.

\section{Calibration of driving amplitude and photon-assisted tunneling}\label{suppsec3}

\begin{figure}[h]
\centering
\includegraphics[width=0.9\textwidth]{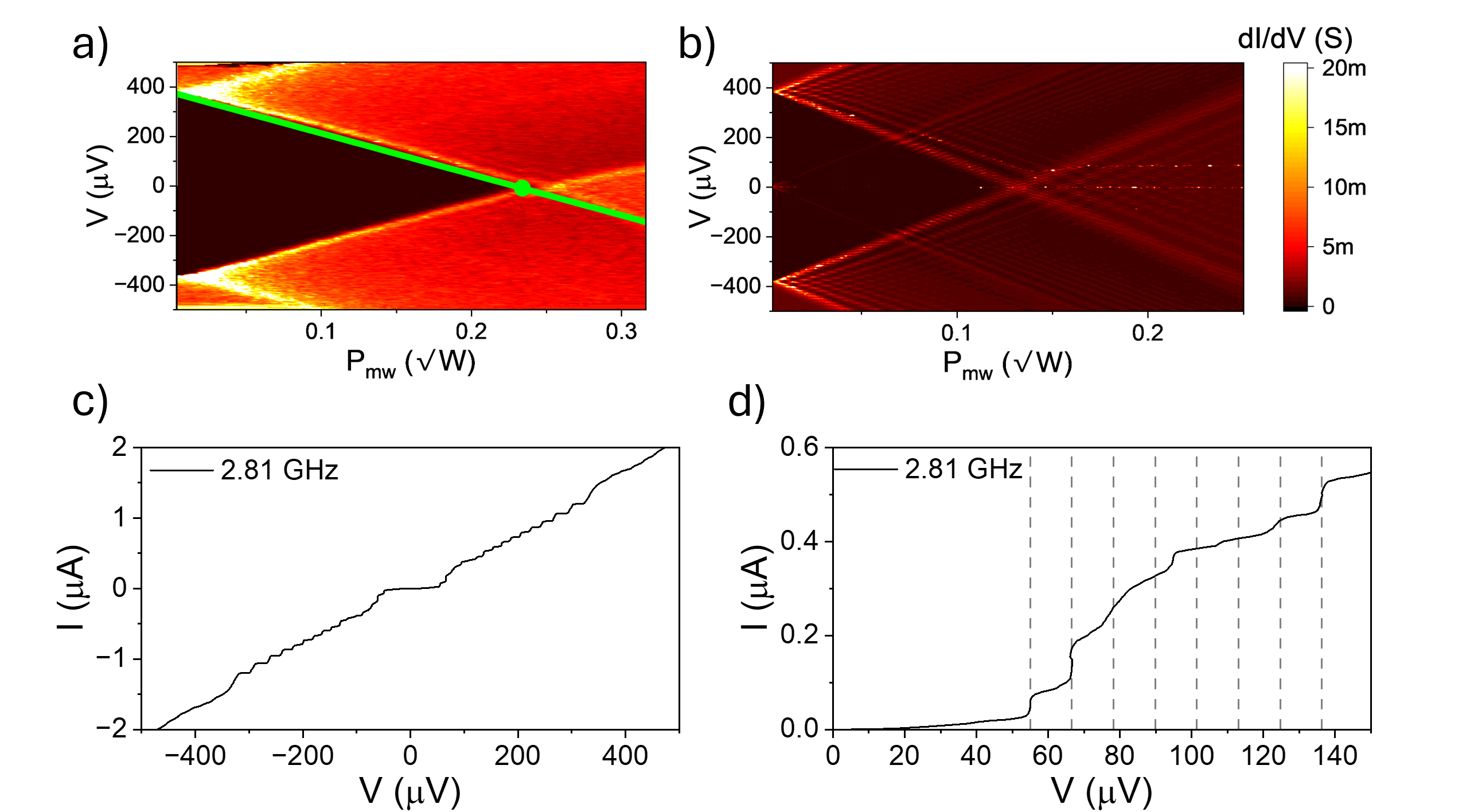}
\caption{\textbf{Evolution of the subgap region with microwave drive power.}. 
\textbf{a}, \textbf{b}, Differential conductance map as a function of bias voltage $V$ and and the square root of microwave AC power $\sqrt{P_{\mathrm{mw}}}$  recorded at $f = 1.405\,\mathrm{GHz}$ ($B = 17\,\mathrm{mT}$, \textbf{a}) and $f = 2.81\,\mathrm{GHz}$ ($B = 0\,\mathrm{mT}$, \textbf{b}). The green curve follow the linear dependence between $V_{\mathrm{subgap}}$ and $\sqrt{P_{\mathrm{mw}}}$.
\textbf{c}, $I\text{--}V$ characteristic measured at at $P_{\mathrm{mw}}\approx19 \,\mathrm{dBm}$, $f = 2.810\,\mathrm{GHz}$ ($B = 0\,\mathrm{mT}$)
showing rich photon-assisted tunneling plateaus. 
\textbf{d}, Close-up view of (\textbf{c}) showing discrete tunneling steps separated by $\Delta V\simeq hf/e=11.62\,\mu\mathrm{V}$. 
Data in \textbf{a}, \textbf{c} and \textbf{d} were measured on Device 1 (main text sample), whereas \textbf{b} were acquired on an additional device (Device 3) with area $11.6\text{ }\mu\text{m}^2$ and $R_N\approx700\text{ }\Omega$.}
\label{figS3}
\end{figure}

The effective microwave amplitude $V_{\mathrm{ac}}$ delivered to the tunnel junction is calibrated using the power-dependent suppression of the superconducting subgap region. According to the theory of photon-assisted tunneling, as $V_{\mathrm{ac}}$ increases, the effective subgap voltage window $V_{\mathrm{subgap}}$ shrinks linearly until closing completely at a threshold amplitude $V_{\mathrm{ac}}^* = 2\Delta/e$, following the relation:
\begin{equation}
V_{\mathrm{subgap}} = \frac{2\Delta}{e} \left(1 - \frac{e V_{\mathrm{ac}}}{2\Delta}\right).
\label{eqV}
\end{equation}

Experimentally, the output power $P_{\mathrm{mw}}$ emitted by the microwave signal generator is controlled directly. This output power relates quadratically to $V_{\mathrm{ac}}$ via $P_{\mathrm{mw}} = V_{\mathrm{ac}}^2 / Z_{\mathrm{load}}$, where $Z_{\mathrm{load}}$ is an effective load impedance accounting for line attenuation, cryogenic filtering, antenna-to-junction coupling and antenna efficiency. The differential conductance maps in Supplementary Fig.~\ref{figS3}a,b display a linear reduction of $V_{\mathrm{subgap}}$ as a function of $\sqrt{P_{\mathrm{mw}}}$,
\begin{equation}
V_{\mathrm{subgap}} = \frac{2\Delta}{e} \left(1 - \frac{\sqrt{P_{\mathrm{mw}}}}{\sqrt{P_{\mathrm{mw}}^*}}\right),
\end{equation}
confirming both the linear trend predicted by Eq.~\eqref{eqV} and the quadratic power-to-voltage relationship, where $P_{\mathrm{mw}}^* = \frac{(2\Delta/e)^2}{Z_{\mathrm{load}}}$.

From the measured slope of $V_{\mathrm{subgap}}$ versus $\sqrt{P_{\mathrm{mw}}}$ or its intercept with the horizontal axis, we extract the conversion factor between $V_{\mathrm{ac}}$ and $P_{\mathrm{mw}}$, $V_{\mathrm{ac}}(P_{\mathrm{mw}}) = \sqrt{Z_{\mathrm{load}} P_{\mathrm{mw}}}$. This allows us to convert the injected microwave power into the effective junction AC voltage drop normalized to $\Delta/e$,
\begin{equation}
\frac{e V_{\mathrm{ac}}}{\Delta} = 2 \frac{\sqrt{P_{\mathrm{mw}}}}{\sqrt{P_{\mathrm{mw}}^*}},
\end{equation}
as used throughout the main text, with a threshold power calibration constant of $P_{\mathrm{mw}}^* = 0.23\text{ }\mathrm{W}$.

\section{Parametric extrema analysis and waveform asymmetry of the biharmonic drive}\label{suppsec4}

To determine the extrema and optimize the spatio-temporal asymmetry of the biharmonic drive voltage waveform, we consider the normalized parametric function:
\begin{equation}
f(x) = \sin(x) + k \sin\left(2x + \frac{\pi}{2}\right), \quad x \in \mathbb{R}, \; k > 0,
\end{equation}
where $x = 2\pi f t$ and $k = V_2 / V_1$ represents the harmonic amplitude ratio. Using the trigonometric identity $\sin\left(\theta + \frac{\pi}{2}\right) = \cos(\theta)$ alongside the double-angle formula $\cos(2x) = 1 - 2\sin^2(x)$, the waveform simplifies to $f(x) = -2k \sin^2(x) + \sin(x) + k$.

Substituting $t = \sin(x)$ with $t \in [-1, 1]$, the extremum problem reduces to finding the bounds of a quadratic polynomial $g(t) = -2kt^2 + t + k$ defined on a compact domain. Because $k > 0$, the quadratic coefficient $a = -2k < 0$ describes a downward-concave parabola with a vertex located at $t_{\mathrm{v}} = 1/(4k)$. The location of $t_{\mathrm{v}}$ relative to $[-1, 1]$ dictates two distinct parametric regimes:

\textbf{Regime A ($0 < k < 1/4$):} Here, $t_{\mathrm{v}} > 1$, placing the vertex outside the physical domain $[-1, 1]$. Consequently, $g(t)$ is strictly monotonic increasing over $[-1, 1]$, yielding a global maximum $\max f = g(1) = 1 - k$ (at $\sin x = 1$) and a global minimum $\min f = g(-1) = -1 - k$ (at $\sin x = -1$).

\textbf{Regime B ($k \ge 1/4$):} Here, $0 < t_{\mathrm{v}} \le 1$, placing the vertex inside $[-1, 1]$. The global maximum corresponds to the vertex value $\max f = g(t_{\mathrm{v}}) = k + 1/(8k) = (8k^2 + 1)/(8k)$. Comparing boundary values ($g(-1) = -1 - k < g(1) = 1 - k$), the global minimum remains localized at $\sin x = -1$, yielding $\min f = -1 - k$.

To evaluate the waveform asymmetry driving non-reciprocal transport, we introduce the dimensionless asymmetry factor $\mathcal{A}(k)$, defined as:
\begin{equation}
\mathcal{A}(k) = \frac{\max f + \min f}{\max f - \min f}.
\end{equation}

Evaluating $\mathcal{A}(k)$ across both parametric regimes gives:
\begin{equation}
\mathcal{A}(k) = 
\begin{cases} 
-k, & \text{for } 0 < k < \dfrac{1}{4}, \\[2ex]
\dfrac{1 - 8k}{(4k + 1)^2}, & \text{for } k \ge \dfrac{1}{4}.
\end{cases}
\label{eq:asymmetry_factor}
\end{equation}

To identify the harmonic amplitude ratio that maximizes drive asymmetry, we differentiate $\mathcal{A}(k)$ with respect to $k$ for $k \ge 1/4$:
\begin{equation}
\frac{d\mathcal{A}}{dk} = \frac{-8(2 - 4k)}{(4k + 1)^3},
\end{equation}
which vanishes strictly at $k = 0.5$. At this optimal setting ($k = V_2 / V_1 = 0.5$, or $V_1 = 2V_2$ as implemented throughout the main text), the asymmetry factor reaches its maximum absolute value $|\mathcal{A}| = 1/3 \approx 33.3\%$.

\section{Extended Tien-Gordon model for biharmonic photon-assisted tunneling}\label{suppsec5}
\begin{figure}[h]
\centering

\includegraphics[width=0.5\textwidth]{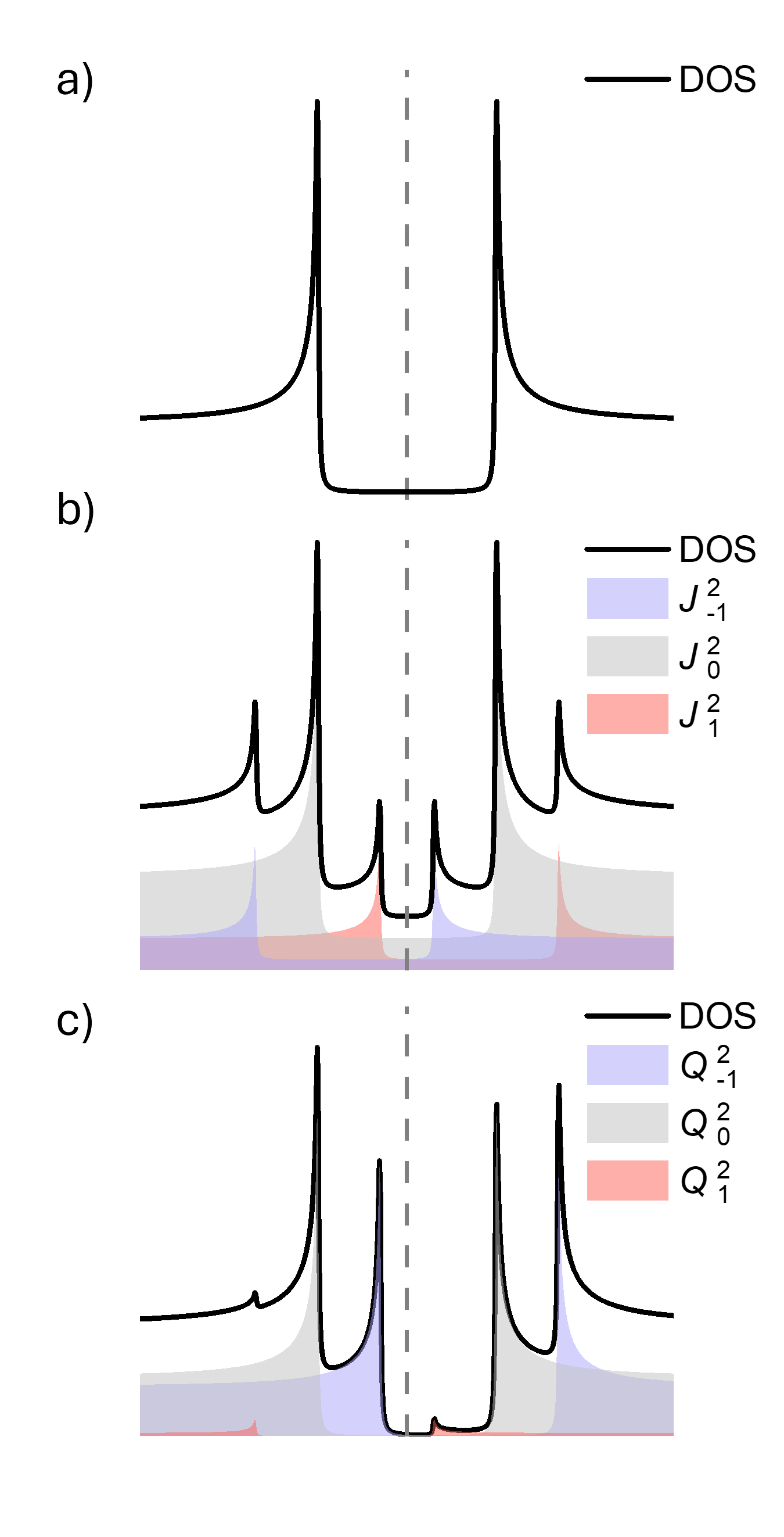}
\caption{\textbf{Comparison between photon excited density of states.} 
\textbf{a}, Unperturbed BCS density of states $D_0(E)$. 
\textbf{b}, Symmetric density of states under single-tone microwave driving and evaluated from the first orders in the Bessel functions $J_n$ at $\alpha_1=1$.
\textbf{c}, Asymmetric density of states under biharmonic driving truncated at low order of $Q_n(\alpha_1=1, \alpha_2=1, \theta=\pi/2)$.
Calculated using Dynes parameter $\Gamma=0.01 \,\Delta$ and microwave frequency $f=0.7 \Delta/h$.}
\label{figS4}
\end{figure}

In an unperturbed SIS junction, the static quasiparticle tunneling current $I_{\mathrm{qp}}^0(V)$ is evaluated through the tunneling integral:
\begin{equation}
I_{\mathrm{qp}}^0(V) = \frac{1}{e R_{\mathrm{N}}} \int_{-\infty}^{+\infty} D_0(E) D_0(E + e V) \left[ n_{\mathrm{F}}(E) - n_{\mathrm{F}}(E+eV) \right] dE,
\end{equation}
where $V$ is the applied voltage, $n_{\mathrm{F}}(E) = \left[1 + \exp(E / k_{\mathrm{B}} T)\right]^{-1}$ is the Fermi-Dirac distribution function accounting for thermal excitations, and $D_0(E) = \operatorname{Re}\left[\frac{E+i\Gamma}{\sqrt{(E+i\Gamma)^2-\Delta^2}}\right]$ is the normalized BCS quasiparticle density of states of the two superconducting electrodes.

Single-quasiparticle tunneling under monochromatic microwave irradiation is modified by introducing transition rates shifted in energy by $\pm n h f$ via Fermi's golden rule, accounting for the absorption or emission of $n$ photons of frequency $f$.
In the classical Tien-Gordon framework~\cite{tien_multiphoton_1963}, these additional transition rates effectively modify the voltage biased unperturbed BCS density of states $D_0(E+eV)$. For a single-tone AC drive $V_{\mathrm{ac}}(t) = V_1 \sin(2 \pi f t)$, the photon-dressed density of states takes the form:
\begin{equation}
D(E+eV) = \sum_{n=-\infty}^{+\infty} J_n^2(\alpha_1) D_0(E+eV + n h f),
\label{eq:TG_monochromatic_DOS}
\end{equation}
where $\alpha_1 = e V_1 / h f$ and $J_n$ is the $n$-th order Bessel function of the first kind. Consequently, the resulting quasiparticle tunneling current is given by the Tien-Gordon formula:
\begin{equation}
I_{\mathrm{qp}}(V) = \sum_{n=-\infty}^{+\infty} p_n I_{\mathrm{qp}}^0\left(V + \frac{n h f}{e}\right),
\label{eqIqp}
\end{equation}
where $p_n = J_n^2(\alpha_1)$ is the probability of exchanging $n$ photons during the tunneling process. Because $J_n^2(\alpha_1) = J_{-n}^2(\alpha_1)$, single-tone driving preserves equal probabilities $p_n = p_{-n}$ for photon absorption ($+n$) and emission ($-n$). Combined with the anti-symmetry of the unperturbed $I\text{--}V$ characteristic $I_{\mathrm{qp}}^0(-V) = -I_{\mathrm{qp}}^0(V)$, this symmetry preserves exact transport reciprocity $I_{\mathrm{qp}}(-V) = -I_{\mathrm{qp}}(V)$.

A biharmonic signal of the form 
\begin{equation}
V_{\mathrm{drive}}(t) = V_1 \sin(2\pi f t) + V_2 \sin(4\pi f t + \theta)
\label{eqBih}
\end{equation}
instead generates a more complex, phase-sensitive density of states structure, weighted by combinations of Bessel functions that depend explicitly on the phase shift $\theta$ between the two harmonics:
\begin{equation}
D(E+eV)=\sum_{n=-\infty}^{+\infty}p_n(V_1,V_2,\theta)D_0(E+eV+nhf)
\label{eqDOS}
\end{equation}
where the multi-photon exchange probability $p_n(V_1, V_2, \theta)$ is given by:
\begin{equation}
p_n(V_1, V_2, \theta) = \left| Q_n(V_1, V_2, \theta) \right|^2 = \left| \sum_{m=-\infty}^{+\infty} J_{n-2m}(\alpha_1) J_m(\alpha_2) e^{i m (\theta - \pi/2)} \right|^2,
\label{eqPn}
\end{equation}
with dimensionless drive parameters $\alpha_1 = e V_1 / h f$ and $\alpha_2 = e V_2 / 2 h f$.

The fundamental difference relative to the single-tone case is that $p_n(V_1, V_2, \theta) \neq p_{-n}(V_1, V_2, \theta)$ for $\theta \neq 0, \pi$. Consequently, the $\pm n$ energy shift rates no longer contribute symmetrically to forward and reverse bias. This unbalance selectively promotes single-quasiparticle tunneling in one bias direction while suppressing it in the opposite direction, breaking the transport symmetry such that $I_{\mathrm{qp}}(-V) \neq -I_{\mathrm{qp}}(V)$. This microscopic asymmetry is directly manifested in the photon-dressed density of states (Supplementary Fig.~\ref{figS4}c), giving rise to a fully non-reciprocal $I\text{--}V$ characteristic.

\subsection{Derivation of biharmonic multi-photon exchange amplitudes}

Consider a single-particle wavefunction $\psi(\mathbf{r}, t) = f(\mathbf{r})\chi(t)$ satisfying the unperturbed single-particle Hamiltonian $H_0$. In the presence of a biharmonic voltage drive of Eq.~\ref{eqBih}, the time-dependent Hamiltonian becomes:
\begin{equation}
H(t) = H_0 + e V_1 \sin(\omega t) + e V_2 \sin(2\omega t + \theta),
\end{equation}
where $\omega = 2\pi f$. The time-dependent phase factor $\chi(t)$ obeys the Schrödinger equation:
\begin{equation}
\chi(t) = \exp\left\{ -\frac{i}{\hbar} \int_0^t \left[ E + e V_1 \sin(\omega t') + e V_2 \sin(2\omega t' + \theta) \right] dt' \right\},
\end{equation}
which evaluates explicitly to:
\begin{equation}
\chi(t) = \exp\left(-\frac{i}{\hbar} E t\right) \exp\left[ \frac{i e V_1}{\hbar\omega} (\cos\omega t - 1) \right] \exp\left[ \frac{i e V_2}{2\hbar\omega} \left( \cos(2\omega t + \theta) - \cos\theta \right) \right].
\end{equation}
Grouping all time-independent phase factors into a constant $A$ (with $|A| = 1$) and defining the dimensionless drive parameters $
\alpha_1 = \frac{e V_1}{\hbar\omega}$, $ \alpha_2 = \frac{e V_2}{2\hbar\omega}$ we obtain:
\begin{equation}
\chi(t) = A e^{-\frac{i}{\hbar} E t} e^{i \alpha_1 \cos\omega t} e^{i \alpha_2 \cos(2\omega t + \theta)}.
\end{equation}

Applying the Jacobi-Anger expansion $e^{i z \cos\phi} = \sum_{k=-\infty}^{+\infty} i^k J_k(z) e^{i k \phi}$ to each drive tone yields:
\begin{align}
e^{i \alpha_1 \cos\omega t} &= \sum_{k=-\infty}^{+\infty} i^k J_k(\alpha_1) e^{i k \omega t}, \\
e^{i \alpha_2 \cos(2\omega t + \theta)} &= \sum_{m=-\infty}^{+\infty} i^m J_m(\alpha_2) e^{i m (2\omega t + \theta)}.
\end{align}
Multiplying these two series, the time-dependent wavefunction factor becomes:
\begin{equation}
\chi(t) = A e^{-\frac{i}{\hbar} E t} \sum_{k=-\infty}^{+\infty} \sum_{m=-\infty}^{+\infty} i^{k+m} J_k(\alpha_1) J_m(\alpha_2) e^{i(k + 2m)\omega t} e^{i m \theta}.
\end{equation}

Substituting $n = k + 2m$ (so that $k = n - 2m$), the double summation transforms to:
\begin{equation}
\chi(t) = A e^{-\frac{i}{\hbar} E t} \sum_{n=-\infty}^{+\infty} \sum_{m=-\infty}^{+\infty} i^{n - 2m + m} J_{n-2m}(\alpha_1) J_m(\alpha_2) e^{i n \omega t} e^{i m \theta}.
\end{equation}
Noting that $i^{-2m+m}=i^{-m} = e^{-i m \pi / 2}$, we pull $i^n$ outside the inner summation:
\begin{equation}
\chi(t) = A e^{-\frac{i}{\hbar} E t} \sum_{n=-\infty}^{+\infty} i^n \left[ \sum_{m=-\infty}^{+\infty} J_{n-2m}(\alpha_1) J_m(\alpha_2) e^{i m (\theta - \pi/2)} \right] e^{i n \omega t}.
\end{equation}

We identify the bracketed term as the generalized multi-photon amplitude $Q_n(\alpha_1, \alpha_2, \theta)$:
\begin{equation}
Q_n(\alpha_1, \alpha_2, \theta) = \sum_{m=-\infty}^{+\infty} J_{n-2m}(\alpha_1) J_m(\alpha_2) e^{i m (\theta - \pi/2)}.
\label{eq:Q_n_definition}
\end{equation}
Thus, the time-dependent wavefunction simplifies to:
\begin{equation}
\chi(t) = A e^{-\frac{i}{\hbar} E t} \sum_{n=-\infty}^{+\infty} B_n e^{i n \omega t},
\end{equation}
where $B_n \equiv i^n Q_n(\alpha_1, \alpha_2, \theta)$. In the single-tone limit ($\alpha_2 = 0$), $J_m(0) = \delta_{m,0}$, reducing $B_n$ to the standard Tien-Gordon amplitude $i^n J_n(\alpha_1)$.

\subsection{Decomposition into incoherent and quantum interference density of states}

The probability $p_n(V_1, V_2, \theta)$ of exchanging a net energy $n\hbar\omega$ during tunneling is given by $p_n = |Q_n|^2 = Q_n Q_n^*$:
\begin{equation}
p_n(V_1, V_2, \theta) = \sum_{m=-\infty}^{+\infty} \sum_{m'=-\infty}^{+\infty} J_{n-2m}(\alpha_1) J_m(\alpha_2) J_{n-2m'}(\alpha_1) J_{m'}(\alpha_2) e^{i(m - m')(\theta - \pi/2)}.
\label{eq:pn_double_sum}
\end{equation}

Splitting this double sum into diagonal ($m = m'$) and off-diagonal ($m \neq m'$) terms isolates the phase-independent and phase-dependent channels:
\begin{equation}
p_n = \sum_{m=-\infty}^{+\infty} J_{n-2m}^2(\alpha_1) J_m^2(\alpha_2) + \sum_{m \neq m'} J_{n-2m}(\alpha_1) J_m(\alpha_2) J_{n-2m'}(\alpha_1) J_{m'}(\alpha_2) e^{i(m - m')(\theta - \pi/2)}.
\end{equation}

For every pair $(m, m')$, its conjugate partner $(m', m)$ combines via Euler's formula $e^{i\phi} + e^{-i\phi} = 2\cos\phi$. Restricting the off-diagonal sum to unordered pairs $m < m'$ yields:
\begin{equation}
\begin{aligned}
p_n(V_1, V_2, \theta) = &\sum_{m=-\infty}^{+\infty} J_{n-2m}^2(\alpha_1) J_m^2(\alpha_2) \\
&+ 2 \sum_{m < m'} J_{n-2m}(\alpha_1) J_m(\alpha_2) J_{n-2m'}(\alpha_1) J_{m'}(\alpha_2) \cos\!\left[(m' - m)(\theta - \pi/2)\right].
\end{aligned}
\label{eq:pn_final_split}
\end{equation}

The biharmonic photon-dressed density of states $D(E) = \sum_n p_n D_0(E + n\hbar\omega)$ naturally decomposes into an incoherent background $D_{\mathrm{inc}}(E)$ and a phase-sensitive quantum interference term $D_{\mathrm{int}}(E, \theta)$:
\begin{equation}
D(E) = D_{\mathrm{inc}}(E) + D_{\mathrm{int}}(E, \theta),
\end{equation}
where:
\begin{align}
D_{\mathrm{inc}}(E) &= \sum_{k=-\infty}^{+\infty} \sum_{m=-\infty}^{+\infty} J_k^2(\alpha_1) J_m^2(\alpha_2) D_0(E + k\hbar\omega + 2m\hbar\omega), \label{eq:D_inc} \\
D_{\mathrm{int}}(E, \theta) &= 2\sum_{n=-\infty}^{+\infty} \sum_{m < m'} J_{n-2m}(\alpha_1) J_m(\alpha_2) J_{n-2m'}(\alpha_1) J_{m'}(\alpha_2) \cos\!\left[(m' - m)(\theta - \pi/2)\right] D_0(E + n\hbar\omega). \label{eq:D_int}
\end{align}

The incoherent term $D_{\mathrm{inc}}(E)$ represents the classical addition of independent photon absorption/emission events at frequencies $f$ and $2f$. Conversely, $D_{\mathrm{int}}(E, \theta)$ encapsulates quantum interference between different multi-photon pathways leading to the same net sideband $n\hbar\omega$.

\subsection{Tien-Gordon current calculation and Gaussian voltage broadening}

Inserting the exspression of $p_n(V_1,V_2,\theta)$ into the quasiparticle tunneling current Eq.\ref{eqIqp}~ or Eq.~3 of the main text allows to evaluate the diode $I\text{--}V$ characteristic. A numerical solution is shown as the black curve in Supplementary Fig.~\ref{fig:Tien-Gordon_smoothen}.

\begin{figure}[h]
\centering
\includegraphics[width=0.95\textwidth]{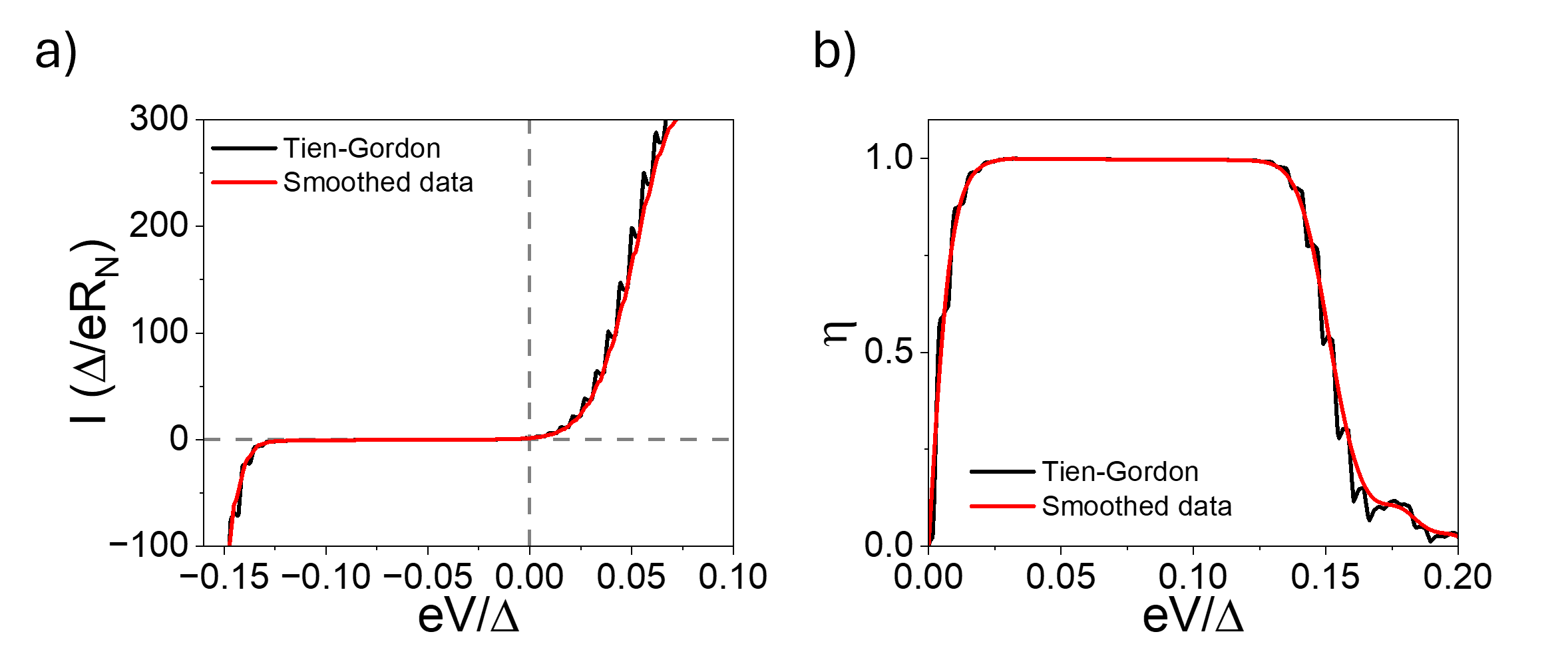}
\caption{\textbf{Biharmonic $I\text{--}V$ characteristics and efficiency evaluated from extended Tien-Gordon theory.} 
\textbf{a}, \textbf{b}, Quasiparticle current (\textbf{a}) and diode efficiency $\eta_{\mathrm{QP}}$ (\textbf{b}) as a function of bias voltage $V$ computed using the biharmonic Tien-Gordon model (Eqs.~\eqref{eqIqp} and \eqref{eq:I_broaden}). Black curves correspond to the raw theoretical model; red curves incorporate Gaussian voltage fluctuations ($\sigma = 3\,\mu\mathrm{V}$). Parameters: $\Delta = 190\,\mu\mathrm{eV}$, $T = 200\,\mathrm{mK}$, $R_{\mathrm{N}} = 155\,\Omega$, $f = 1.4\,\mathrm{GHz}$, $V_1 = 291\,\mathrm{mV}$, and $V_2 =58\,\mathrm{mV}$.}
\label{fig:Tien-Gordon_smoothen}
\end{figure}

To account for thermal noise and low-frequency voltage fluctuations present in the experimental setup, we convolute $I(V)$ with a Gaussian voltage distribution $G_\sigma(V)$:
\begin{equation}
\label{eq:I_broaden}
I_{\mathrm{b}}(V) = \int_{-\infty}^{+\infty} I(V') G_\sigma(V - V') dV', \qquad G_\sigma(V) = \frac{1}{\sqrt{2\pi}\sigma} \exp\left(-\frac{V^2}{2\sigma^2}\right),
\end{equation}
where $\sigma$ represents the standard deviation of the effective voltage noise. As shown in Supplementary Fig.~\ref{fig:Tien-Gordon_smoothen}, introducing $\sigma = 3\,\mu\mathrm{V}$ smooths the sharp photon-assisted tunneling steps, producing quantitative agreement with measured experimental traces.

\subsection{Microscopic mechanism of multi-photon quantum interference}

The microscopic origin of non-reciprocal transport lies in phase-dependent quantum interference among competing transition pathways, as described by Eqs.~\eqref{eqDOS}~\eqref{eqPn} and \eqref{eq:D_int}. Index $n$ labels the final energy sideband $n h f$, whereas index $m$ denotes individual pathways involving the exchange of $m$ photons of frequency $2f$ and $n - 2m$ photons of frequency $f$.

\begin{figure}[h]
\centering
\includegraphics[width=0.85\textwidth]{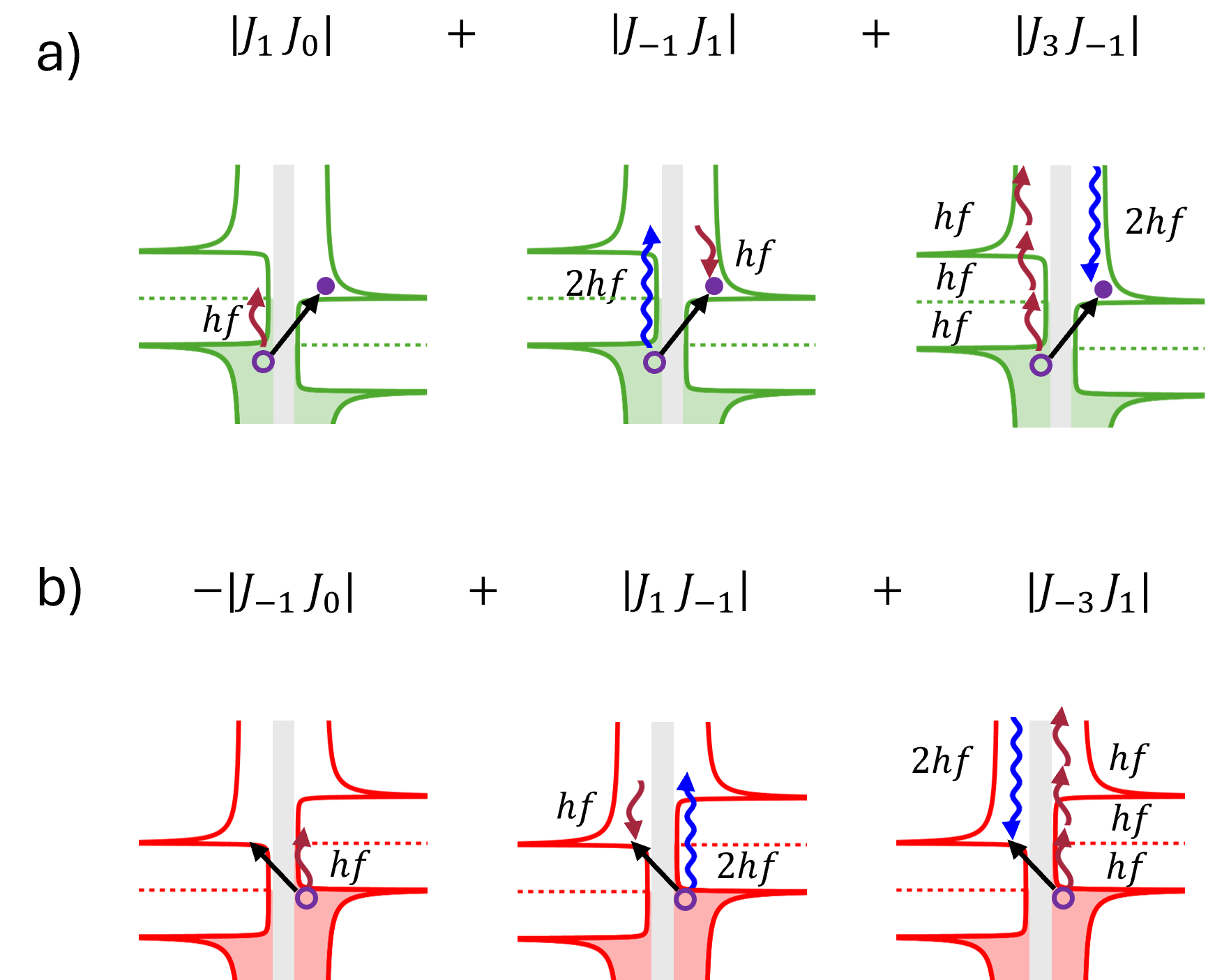}
\caption{\textbf{Quantum interference pathways for multi-photon sidebands $\pm n h f$.} 
\textbf{a}, \textbf{b}, Multi-photon transition amplitudes $Q_n$ for $n = +1$ (\textbf{a}) and $n = -1$ (\textbf{b}) with $m = 0, \pm 1$, illustrating constructive (\textbf{a}) and destructive (\textbf{b}) quantum interference at $\theta = -\pi/2$. Red and blue arrows represent photons of frequency $f$ and $2f$, respectively; black arrows indicate the net transition process, with upward and downward arrows denoting absorption ($m > 0$) and emission ($m < 0$).}
\label{figS5}
\end{figure}

For $\alpha_1, \alpha_2 \ll 1$, we may restrict $Q_n$ to the dominant lowest-order pathways ($m = 0, \pm 1$). At phase setting $\theta = -\pi/2$, Eq.~\eqref{eq:Q_n_definition} expands to:
\begin{equation}
Q_n \approx J_{n+2}(\alpha_1) J_{-1}(\alpha_2) e^{i\pi} + J_n(\alpha_1) J_0(\alpha_2) + J_{n-2}(\alpha_1) J_1(\alpha_2) e^{-i\pi}.
\end{equation}
Using $e^{\pm i\pi} = -1$ and $J_{-k}(z) = (-1)^k J_k(z)$, we evaluate the forward ($n = +1$) and reverse ($n = -1$) transition amplitudes:

For $n = +1$ (forward bias, positive sideband):
\begin{equation}
Q_{+1} \approx J_3(\alpha_1) J_1(\alpha_2) + J_1(\alpha_1) J_0(\alpha_2) + J_1(\alpha_1) J_1(\alpha_2).
\end{equation}
All partial amplitudes share identical positive signs and interfere \textbf{constructively}, strongly enhancing forward quasiparticle tunneling.

For $n = -1$ (reverse bias, negative sideband):
\begin{equation}
Q_{-1} \approx J_1(\alpha_1) J_1(\alpha_2) - J_1(\alpha_1) J_0(\alpha_2) + J_3(\alpha_1) J_1(\alpha_2).
\end{equation}
The partial amplitudes oppose each other in sign and interfere \textbf{destructively}, suppressing reverse quasiparticle tunneling. The photon-assisted tunneling dynamics for these two cases are illustrated in Supplementary Fig.~\ref{figS5}.

\section{Generalized framework for turn-on ($V_{\mathrm{on}}$) and breakdown ($V_{\mathrm{BD}}$) voltages under biharmonic drive}\label{suppsec6}

In a Josephson tunnel junction operating in the quasiparticle regime, the unperturbed insulating state persists as long as the bias potential across the barrier remains within the superconducting energy gap, $e |V| < 2\Delta$. Under a general time-dependent AC voltage bias $V_{\mathrm{ac}}(t) = V_1 \sin(\omega t) + V_2 \sin(2\omega t + \theta)$ superimposed on a static DC bias $V_{\mathrm{dc}}$, the total instantaneous potential across the junction is $V(t) = V_{\mathrm{dc}} + V_{\mathrm{ac}}(t)$.

In the quasistatic limit ($\hbar\omega \ll \Delta$), macroscopically observable quasiparticle conduction is triggered whenever the peak of the instantaneous total voltage $V(t)$ crosses the continuum gap boundary $\pm 2\Delta/e$. This defines two distinct voltage thresholds depending on the bias polarity:
\begin{enumerate}
    \item Forward Turn-On Voltage ($V_{\mathrm{on}}$): The minimum positive DC bias ($V_{\mathrm{dc}} > 0$) required for the maximum positive excursion of the total waveform to reach the positive gap edge $+2\Delta/e$:
    \begin{equation}
    V_{\mathrm{on}} + \max_{t} \left[ V_{\mathrm{ac}}(t) \right] = \frac{2\Delta}{e} \implies V_{\mathrm{on}} = \frac{2\Delta}{e} - V_{\mathrm{peak}, +},
    \label{eq:Von_general}
    \end{equation}
    where $V_{\mathrm{peak}, +} \equiv \max_t [V_{\mathrm{ac}}(t)]$ is the positive waveform peak.

    \item Reverse Breakdown Voltage ($V_{\mathrm{BD}}$): The minimum magnitude of negative DC bias ($V_{\mathrm{dc}} < 0$) required for the maximum negative excursion of the total waveform to reach the negative gap edge $-2\Delta/e$:
    \begin{equation}
    -|V_{\mathrm{BD}}| + \min_{t} \left[ V_{\mathrm{ac}}(t) \right] = -\frac{2\Delta}{e} \implies |V_{\mathrm{BD}}| = \frac{2\Delta}{e} - |V_{\mathrm{peak}, -}|,
    \label{eq:Vbr_general}
    \end{equation}
    where $|V_{\mathrm{peak}, -}| \equiv |\min_t [V_{\mathrm{ac}}(t)]|$ is the absolute magnitude of the negative waveform peak.
\end{enumerate}

As established in Section~\ref{suppsec4}, the peak amplitudes $V_{\mathrm{peak}, +}$ and $|V_{\mathrm{peak}, -}|$ are continuously modulated by the relative phase $\theta$ and harmonic ratio $k = V_2/V_1$. At the optimal drive configuration ($k = 0.5$ and $\theta = -\pi/2$), the peak limits reach $V_{\mathrm{peak}, +} = 1.5 V_1$ and $|V_{\mathrm{peak}, -}| = 0.75 V_1$.

Quasiparticle conduction is activated whenever the total instantaneous voltage reaches the gap threshold $2\Delta/e$. Requiring an \textbf{ideal diode condition} with zero turn-on voltage ($V_{\mathrm{on}} = 0$) strictly fixes the fundamental drive amplitude $V_1$:
\begin{equation}
V_{\mathrm{on}} = \frac{2\Delta}{e} - 1.5 V_1 = 0 \implies V_1 = \frac{4}{3} \frac{\Delta}{e} \approx 253.3\,\mu\mathrm{V}.
\label{eq:ideal_V1}
\end{equation}

Inserting Eq.~\eqref{eq:ideal_V1} into the reverse breakdown condition $|V_{\mathrm{BD}}| + |V_{\mathrm{peak}, -}| = 2\Delta/e$ determines the reverse breakdown voltage:
\begin{equation}
|V_{\mathrm{BD}}| = \frac{2\Delta}{e} - 0.75 V_1 = \frac{2\Delta}{e} - 0.75 \left(\frac{4\Delta}{3e}\right) = \frac{\Delta}{e} \approx 190\,\mu\mathrm{V}.
\label{eq:ideal_Vbr}
\end{equation}
In this semiclassical picture, the diode exhibits zero current for $-\Delta/e < V_{\mathrm{dc}} < 0$, a step-like activation at $V_{\mathrm{on}} = 0$, and an abrupt reverse breakdown at $|V_{\mathrm{BD}}| = \Delta/e$.

\section{Effective RSJ model}\label{suppsec7}
\begin{figure}
    \centering
\includegraphics[width=1\textwidth]{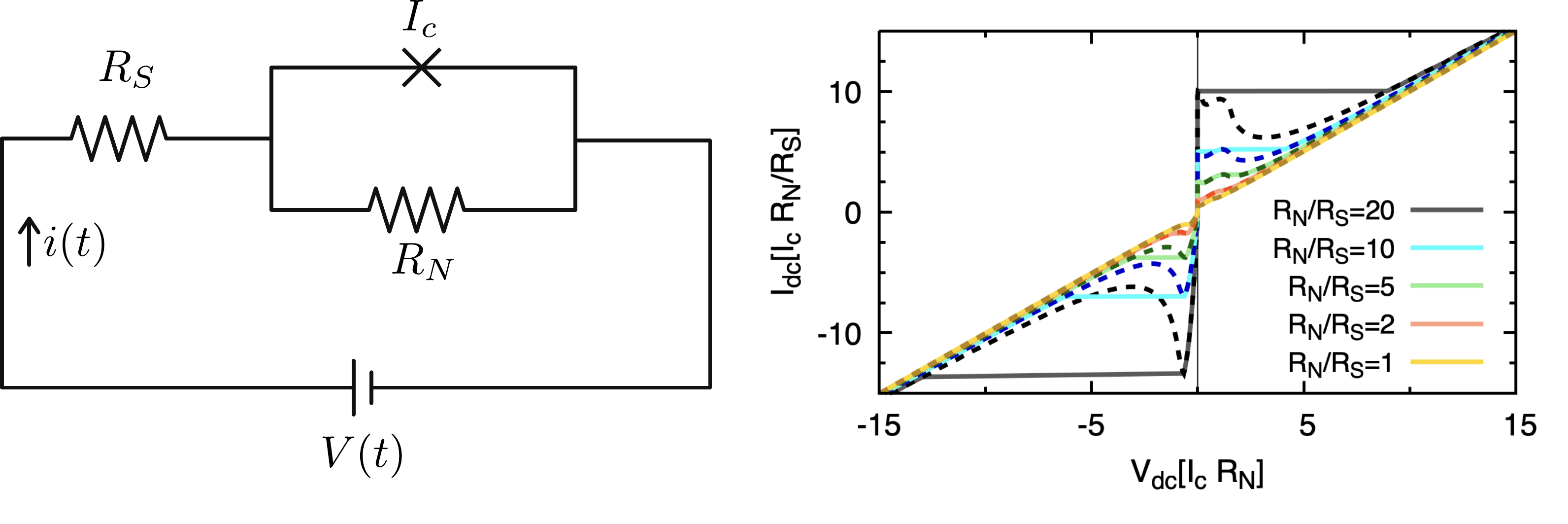}
\caption{Left: Schematic of the modified RSJ circuit. Right panel: simulated $I\text{--}V$ characteristic for a biharmonic drive with $\omega=0.001$, $V_1=2V_2=I_c/3$, $\theta=\pi/2$, $R_N=1$, and $R_s$=2. Here, we have considered $I_c=1$ as the current unit.}
\label{modified_RSJ}
\end{figure}

To describe the dynamics of the system in the superconducting state, we use an adapted Resistively Shunted Junction (RSJ) model, driven by voltage $V(t)$ through a source resistance $R_{\mathrm{S}}$ as described in Fig.~\ref{modified_RSJ}.
The phase dynamics of the modified RSJ circuit are described by
\begin{equation}
\frac{\hbar}{2e}\frac{\dot{\phi}}{R_N} = -I_c\cos\phi+i(t).
\end{equation}
The circuit current is given by
\begin{equation}
i(t)=\frac{V(t)-(\hbar/2e)\dot{\phi}}{R_S}.
\end{equation}
The voltage drop across the Josephson junction is given by
\begin{equation}
V_{dc} = \frac{\hbar}{2e}\left\langle \dot{\phi} \right\rangle,
\end{equation}
where the brackets indicate the time average and
\begin{equation}
I_{dc} = \left\langle i(t) \right\rangle.
\end{equation}

Results are shown in the right panel of fig.~\ref{modified_RSJ}, where the system features a region with intermediate resistance (smaller than $R_N$) for low voltage in only one of the bias direction. In this region, the phase is well-localized some part of the time period, inducing dissipation only for some part.

At the transition to the running-phase state (see dashed lines), the average voltage developed across the junction can increase faster than the applied bias, thereby reducing the voltage drop across $R_S$ and producing a branch with negative differential conductance. Such a branch is unstable in the physical circuit, causing the system to switch between the neighboring stable solutions rather than continuously following the negative-slope branch, illustrated by the solid lines.

\end{document}